\documentclass[aps,prb,twocolumn,groupedaddress]{revtex4-1}

\usepackage{graphicx,epstopdf,siunitx,braket,xspace,amssymb,dsfont,bm,amsmath,float}
\usepackage{adjustbox}
\usepackage{booktabs}
\usepackage[utf8]{inputenc}

\newcommand{\mc}[1]{\ensuremath{\mathcal{#1}}}
\newcommand{\mr}[1]{\ensuremath{\mathrm{#1}}}
\newcommand{\je}[1]{\ensuremath{J_{#1}(\epsilon_{#1})}\xspace}
\newcommand{\dbz}[0]{\ensuremath{b}\xspace}
\newcommand{\db}[1]{\ensuremath{b_{#1}}\xspace}
\newcommand{\bg}[0]{\ensuremath{B_\mr{G}}\xspace}
\newcommand{\eps}[0]{\ensuremath{\epsilon}\xspace}
\newcommand{\fid}[0]{\ensuremath{\mc{F}}\xspace}
\newcommand{\infid}[0]{\ensuremath{\mc{I}}\xspace}
\newcommand{\leak}[0]{\ensuremath{\mc{L}}\xspace}
\newcommand{\lc}[0]{\ensuremath{\mc{L}_{\mr{c}}}\xspace}
\newcommand{\li}[0]{\ensuremath{\mc{L}_{\mr{i}}}\xspace}
\newcommand{\ifef}[0]{\ensuremath{\mc{I}_{\mr{f}}}\xspace}
\newcommand{\ifes}[0]{\ensuremath{\mc{I}_{\mr{s}}}\xspace}
\newcommand{\ifb}[0]{\ensuremath{\mc{I}_\mr{b}}\xspace}
\newcommand{\ifu}[0]{\ensuremath{\mc{I}_\mr{u}}\xspace}

\newcommand{\qpef}[0]{\ensuremath{\mc{E}_{\mr{f}}}\xspace}
\newcommand{\qpes}[0]{\ensuremath{\mc{E}_{\mr{s}}}\xspace}
\newcommand{\qpb}[0]{\ensuremath{\mc{E}_\mr{b}}\xspace}

\newcommand{\vecsig}[0]{\ensuremath{\bm{\sigma}}\xspace}

\newcommand{\pz}[0]{\ensuremath{\sigma_z}\xspace}
\newcommand{\eye}[0]{\ensuremath{I}\xspace}
\newcommand{\ec}[0]{\ensuremath{E_\mr{c}}\xspace}
\newcommand{\hc}[0]{\ensuremath{H_\mr{c}}\xspace}
\newcommand{\uc}[0]{\ensuremath{U_\mathrm{c}}\xspace}
\newcommand{\ut}[0]{\ensuremath{U_\mathrm{t}}\xspace}
\newcommand{\uf}[0]{\ensuremath{U_\mathrm{f}}\xspace}
\newcommand{\fs}[0]{\ensuremath{f_{\mr{s}}}\xspace}
\newcommand{\matdist}[0]{\ensuremath{\mathbf{\Delta}}\xspace}

\newcommand{\reftab}[1]{Tab.\,\ref{#1}}
\newcommand{\reffig}[1]{Fig.\,\ref{#1}\,}
\newcommand{\refsec}[1]{Appendix \,\ref{#1}}
\newcommand{\sts}[0]{$\mathrm{S\mbox{-}T_0}$\xspace}
\newcommand{\kets}[0]{\ensuremath{\ket{\mr{S}}}\xspace}
\newcommand{\kett}[0]{\ensuremath{\ket{\mr{T_0}}}\xspace}	
\newcommand{\nseg}[0]{\ensuremath{N_{\mr{seg}}}\xspace}

\newcommand{\lind}[0]{\ensuremath{\mathcal{L}}}

\begin{document}

\title{High-fidelity gate set for exchange-coupled singlet-triplet qubits}
\author{Pascal Cerfontaine}
\email[]{pascal.cerfontaine@rwth-aachen.de}
\author{Ren\'e Otten}
\author{M. A. Wolfe}
\author{Patrick Bethke}
\author{Hendrik Bluhm}
\affiliation{JARA-FIT Institute for Quantum Information, Forschungszentrum J\"ulich GmbH and RWTH Aachen University, 52074 Aachen, Germany}


\date{April 28, 2020}

\begin{abstract}
In order to enable semiconductor-based quantum computing with many qubits, issues like residual interqubit coupling and constraints from scalable control hardware need to be tackled to retain the high gate fidelities demonstrated in current single-qubit devices. Here, we focus on two exchange-coupled singlet-triplet spin qubits, considering realistic control hardware as well as Coulomb and exchange coupling that cannot be fully turned off. Using measured noise spectra, we optimize realistic control pulses and show that two-qubit (single-qubit) gate fidelities of 99.90\% ($\ge 99.69\%$) can be reached in GaAs, while 99.99\% ($\ge 99.95\%$) can be achieved in Si.
\end{abstract}

\maketitle


\section{Introduction}

Well-controlled qubit arrays with access to a high-fidelity gate set of single- and two-qubit gates are a key ingredient for building a quantum computer. Electron spin qubits in gate-defined quantum dots, which are among the main contenders for scalable quantum computing, have achieved sufficiently high single-qubit gate fidelities exceeding 99.9\% \cite{Yoneda2017,Dehollain2016} using microwave control of individual spins. In contrast, the best two-qubit gate fidelities as well as single-qubit gates in two-qubit devices have not yet exceeded 98\% \cite{Veldhorst2015,Watson2017,Zajac2017,Huang2018a}, still below the threshold required for fault-tolerant quantum computation. A similar trend is seen in superconducting qubits, where gate fidelities tend to decrease as more qubits are added \cite{Kelly2015,Barends2014,Fried2017}. Thus, it seems necessary to treat the implementation of single- and two-qubit gates simultaneously to address major challenges like residual interqubit coupling.

An alternative to single-spin qubits, known as singlet-triplet (\sts) qubits, uses the $m_s=0$ states of two electron spins. This allows all operations to be achieved with sub-GHz baseband control of exchange interactions \cite{Klinovaja2012,Wardrop2014,Li2012,Levy2002,Mehl2014b}, potentially avoiding hardware challenges when scaling up by eliminating the need for microwaves. Their single-qubit operations have already been demonstrated experimentally in GaAs with \SI{99.5}{\%} fidelity \cite{Cerfontaine2019ex} in close agreement with theoretical predictions of \SI{99.6}{\%} and \SI{99.8}{\%}~\cite{Cerfontaine2019ex, Cerfontaine2014}. Two-qubit gates have only been demonstrated using capacitive (Coulomb) coupling~\cite{Shulman2012,Nichol2017}. The resulting fidelities (70-90\%) are much lower than single-qubit gate fidelities due to the relatively weak Coulomb coupling. A promising alternative is to use the much stronger exchange interaction so that two-qubit gates rely on the same ingredients as single-qubit gates. Hence, comparable fidelities can be expected. 

Early theory works on \sts qubits \cite{Klinovaja2012,Wardrop2014,Li2012,Levy2002,Mehl2014b} considered single- and two-qubit gates, where the interqubit exchange coupling can be fully switched off. More recent work also includes residual interqubit coupling \cite{Buterakos2018, Buterakos2018a}. All of these are based on simplified models well suited for conceptual insight. They consider either Coulomb or exchange coupling but not both at the same time. For a significantly higher level of realism, we now include both coupling mechanisms and other experimentally relevant effects simultaneously. We develop experimentally realistic pulse sequences for single-qubit gates and a CNOT gate, accounting for residual interqubit couplings. Their average gate fidelities range from 99.69\% to 99.90\% in GaAs and from 99.95\% to 99.99\% in Si devices (with vanishing magnetic field noise). With 99.90\% in GaAs and 99.99\% in Si, the CNOT fidelity is one to two orders of magnitude better than any experimental result on spin qubits to date. We investigate how these fidelities scale with the noise strengths and manipulation time, and find straightforward scaling laws for adapting our results to different device parameters.

We incorporate all effects relevant to a realistic setting in a way that has been shown to predict achievable gate infidelities within a factor two for single-qubit gates \cite{Cerfontaine2019ex}. Our model includes interqubit Coulomb coupling, the finite dynamic range of interqubit exchange coupling, electric and magnetic noise with realistic noise spectra, finite pulse rise times and other hardware constraints. We use a simulation-based approach, constraining our numerical search for pulse sequences only to the extent imposed by hardware limitations. Since this method can be expected to yield the best possible fidelities for the model at hand, we expect that such a comprehensive treatment is of high value for predicting the performance of specific quantum-computing platforms. Because our results are directly applicable to current experiments, they provide a complete recipe for high-fidelity control of \sts qubits.

\section{Qubit model}
\begin{figure}[ht!]
    \includegraphics{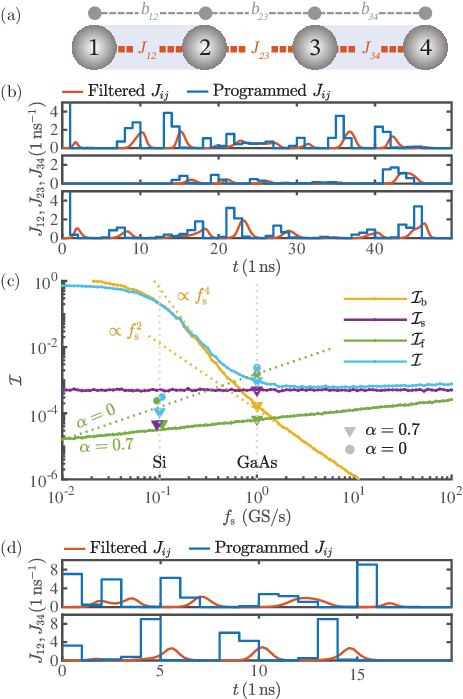}
    \caption{(color online) (a) Quadruple quantum dot configuration forming two \sts qubits with the local exchange interactions $J_{12}$ and $J_{34}$, non-local exchange $J_{23}$, and magnetic field gradients $\db{ij}$. (b) CNOT pulse $J_{ij}(t)$ with $\db{12} = -\db{34} = \SI{1}{ns^{-1}}$ and $\db{23} = \SI{7}{ns^{-1}}$. Blue traces show the sample values to be programmed to the AWG with \SI{1}{ns} resolution. Red traces are convoluted with a measured impulse response, as seen by the qubit. The pulse was optimized for GaAs with $\ec = 0$ and has a fidelity of $\SI{99.90}{\%}$ (see \reftab{tab:best_cnot_fid}). (c) Infidelity contributions of the CNOT gate as a function of the AWG's sample rate for $\alpha = 0.7$. We assume the AWG's rise time scales with the sample rate. The circles (triangles) represent results obtained for $\alpha = 0$ ($0.7$) in Si and GaAs, while the dotted lines show the expected scaling. (d) $X_{\pi/2}\otimes\eye$ pulse for compensating residual coupling optimized for $\ec = \SI{350}{\micro eV}$ and $J_{23} = \SI{0.005}{ns^{-1}}$ with $\fid = \SI{99.69}{\%}$ in GaAs.}
    \label{fig:main}
\end{figure}

We consider two qubits encoded in four linearly adjacent quantum dots in a semiconductor heterostructure (see \reffig{fig:main}(a)). During qubit manipulation, each double dot is tuned in the $(1,1)$ charge state, where $(n,m)$ represents the number of electrons in dots 1 and 2 or dots 3 and 4. Metal top gates control the exchange interaction $J_{ij}$ between two adjacent dots ($i \in \{1,2,3\}, j = i + 1$)  by either changing the detuning voltage $\eps_{ij}$ (affecting the energy difference between dots $i$ and $j$) \cite{Petta2005,Dial2013} or by direct control of the tunnel barrier \cite{Reed2016,Martins2016}. We focus on the widely used detuning since it provides a good on-to-off ratio of the exchange coupling and the effect of noise is better understood. Adapting our approach to barrier control could yield even higher fidelities due to the observed larger number of coherent oscillations \cite{Martins2016}.

The computational subspace is spanned by the $m_s = 0$ singlet and triplet states of the double dots $1,2$ and $3,4$. We use the basis $\ket{00}=\ket{\uparrow \downarrow \uparrow \downarrow}$, $\ket{01}=\ket{\uparrow \downarrow \downarrow \uparrow}$, $\ket{10}=\ket{\downarrow \uparrow \uparrow \downarrow}$, and $\ket{11}=\ket{\downarrow \uparrow \downarrow \uparrow}$. Two other $m_s$~=~0 states, $\ket{\downarrow \downarrow \uparrow \uparrow}$ and $\ket{\uparrow \uparrow \downarrow \downarrow}$, are dynamically accessible via the intermediate exchange $J_{23}$. Any occupation of these leakage states after a gate must be carefully avoided. Leakage into other spin states has been experimentally shown to be small ($\sim \SI{1e-3}{}$ \cite{Cerfontaine2019ex}) due to the Zeeman splitting from an externally applied magnetic field of \SI{500}{mT}. 

Furthermore, each spin experiences a different constant magnetic field $B_i$ typically realized with micromagnets \cite{Wu2014}, gate-voltage-tuning of the electron g-factor \cite{Veldhorst2014} or via dynamic nuclear polarization (DNP) \cite{Bluhm2010}. The complete Hamiltonian describing the four spins in terms of the mean magnetic field $\bg=\frac{1}{4}\sum_{i=1}^4B_i$ and the average magnetic field gradients across two adjacent dots, $\db{ij}=B_j-B_i$ is given by
\begin{align}
\label{eq:hj}
	H = & \sum_{i=1}^3 \frac{J_{i,i+1}}{4}\vecsig ^{(i)} \cdot \vecsig ^{(i+1)} + \frac{1}{2}  \bg  \sum_{i=1}^4 \pz^{(i)} \nonumber \\
	+ & \frac{\db{12}}{8} [-3\pz^{(1)} + \pz^{(2)} + \pz^{(3)} + \pz^{(4)} ] \nonumber \\
	+ & \frac{\db{23}}{4} [-\pz^{(1)} - \pz^{(2)} + \pz^{(3)} + \pz^{(4)} ] \nonumber \\ 
	+ &\frac{\db{34}}{8}  [-\pz^{(1)} - \pz^{(2)} - \pz^{(3)} +3 \pz^{(4)} ], 
\end{align}
where $\sigma^{(i)}$ acts on the spin in quantum dot $i$ and all prefactors have units of angular frequency ($\hbar = 1$ in the Schrödinger equation). We also include capacitive coupling between the qubits by adding the empirical model 
\begin{align}
\label{eq:hc}
    \hc = \ec \frac{\partial J_{12}}{\partial\eps_{12}}\frac{\partial J_{34}}{\partial\eps_{34}}(\eye-\sigma_z)\otimes(\eye-\sigma_z)/4
\end{align}
written in the $(\kett,\kets)^{\otimes 2}$ basis \cite{Shulman2012,Buterakos2018}. This model reflects the notion that the two-qubit phase acquired by $\ket{\mr{SS}}$ is proportional to the detuning-dependent admixture of $(0,2)$ charge states in the two hybridized $|S\rangle$ states, which according to first order perturbation theory is given by $\partial J / \partial\eps$ for each qubit. The prefactor \ec is the charge coupling energy corresponding to a full transition from $(1,1)$ to $(0,2)$ in each double dot.

\section{Control model}
For constant $\db{ij} \neq 0$, any target unitary operation \ut in the computational subspace can be generated by manipulating $\epsilon_{ij}$ and thus $J_{ij}$ as a function of time. However, a straightforward implementation of \ut is complicated by the nonlinear and imperfectly known relation $J_{ij}(\epsilon_{ij})$, noise on the qubit control parameters $\epsilon_{ij}$ and $\db{ij}$, experimental constraints, and residual interqubit coupling. We incorporate these effects as follows when searching for pulse sequences $\epsilon_{ij}(t)$ that realize a given target unitary \ut.

In experiments, $\epsilon_{ij}(t)$ is typically controlled by baseband pulses from arbitrary waveform generators (AWGs). Our model includes constraints introduced by the AWG, like bounds, $\eps_\mr{min} \le \eps_{ij}(t) \le \eps_\mr{max}$, and a fixed sample rate \fs. Fixing the sample rate results in a piece-wise constant time trace $\eps_{ij, k}$ with time index $k = 1 \ldots \nseg$, where the number of segments \nseg is related to the total gate time $T = \nseg/\fs$. Furthermore, we consider the AWG's limited bandwidth which leads to a smooth time trace $\epsilon_{ij}(t)$. We obtain $\epsilon_{ij}(t)$ by convoluting $\eps_{ij, k}$ with the impulse response of a typical experimental setup as discussed in \refsec{sec:A}. We fix the last 4 samples of each pulse at $\eps_\mr{min}$ to ensure that transients settle nearly completely and do not lead to significant errors in subsequent quantum operations. The phenomenological relation for tilt-control $J_{ij}(\epsilon_{ij}) = J_0 \exp{(\epsilon_{ij}/\epsilon_0)}$ allows us to obtain $J_{ij}(t)$ using $\epsilon_0$ and $J_0$ given in \reftab{tab:params}. Our model also includes decoherence from $\epsilon_{ij}$ noise, modelled via a full noise spectrum. Furthermore, we consider quasistatic $\db{ij}$ noise, which fluctuates on much slower time scales than the nanosecond gate times   \cite{Reilly2008, Barthel2009}.

\section{Numerics and optimization}
We now outline how we determine $\eps_{ij, k}$ to generate \ut with high fidelity. For given $J_{ij}(t)$ and $\db{ij}$, we approximate the time-dependent Hamiltonian as piece-wise constant, with time steps chosen sufficiently small to cause negligible errors. We compute the matrix exponential of the 6-dimensional $m_s = 0$ subspace by direct diagonalization in each time step and obtain the full unitary operator \uf from $t = 0$ to $T$ as a function of $\eps_{ij, k}$ and \db{ij}.  We define $V_{\mr{c}}$ as the truncation of \uf into the four-dimensional computational subspace to compute coherent leakage \lc as the distance from unitarity, $\lc = 1 - \mathrm{tr}(V_{\mr{c}}^{\dagger}V_{\mr{c}})/4$. We also map $V_{\mr{c}}$ to the closest unitary \uc \cite{Keller1975} in the computational subspace with the same global phase as \ut. This ensures that the distance from \ut, $\matdist = \ut - \uc$ is independent of the dynamics in the leakage subspace and the global phase.

To evaluate the separate effects of the slow noise contributions, we average over a discrete Gaussian distribution of $\db{ij}$ ($\epsilon_{ij}$) with standard deviation $\sigma_{\mr{\db{}}}$ ($\sigma_{\eps}$), and obtain the quantum process \qpb (\qpes) by computing the unitary generated by the Hamiltonian for each noise offset. During gate optimization, we include fast charge noise fluctuations with a white noise spectrum $S_{\epsilon}(f) = S_0$ to obtain the process \qpef in a computationally efficient manner from a Lindblad equation and a Markov approximation. We quantify the decoherence from each noise source by $\infid_{\mr{b}} = 1 - \fid(\uc, \qpb)$, $\infid_{\mr{s}} = 1 - \fid(\uc, \qpes)$ and $\infid_{\mr{f}} = 1 - \fid(\uc, \qpef)$, using the average gate fidelity \fid \cite{Nielsen2002}.

To find parameters $\epsilon_{ij, k}$ and \db{ij} such that $\uc$ implements $\ut$ with minimal decoherence and leakage, we solve the nonlinear optimization problem $\min_{\eps_{ij,k}}|\matdist, \ifb, \ifes, \ifef, \lc|^2$ with the Levenberg-Marquardt algorithm (LMA). To speed up the algorithm's convergence, we use analytic derivatives of $\uc$ and \qpef to efficiently derive all terms in the minimization problem with respect to $\eps_{ij,k}$, as discussed in greater detail in \refsec{sec:B}. The chance to find or approach the best fidelity for our model is improved by starting the optimization from many random seeds. Accurate control of \db{ij} is often not available in experiments, so we fix \db{ij} at a few experimentally feasible values and only optimize $\eps_{ij,k}$. Further information on numerical pulse optimization can be found in \refsec{sec:B} and Refs. \citenum{Khaneja2005,Kuprov2009,Floether2011}. 

After the optimization, more accurate infidelities and leakages are calculated by computing unitaries for a large number of noise realizations, including auto-correlated charge noise with spectra $S_{\epsilon,\alpha}(f) \propto 1/f^\alpha$. Additional to coherent leakage, we also compute the total leakage \leak by averaging \lc over the noise realizations, and define the incoherent leakage $\li = \leak - \lc$.

\section{Parameter values}
We choose the experimental parameters given in \reftab{tab:params} for our optimization, and then generalize our findings to different gate durations and noise strengths as given in \reffig{fig:main}(c) and \reffig{fig:best_cnot_noise_scaling_ec0}. Since charge noise spectra in Si differ greatly between devices \cite{Chan2018, Mi2018, Yoneda2017, Eng2015}, we choose GaAs parameters, which are on the low end of what has been measured in Si. The high-frequency spectrum follows $S_{\epsilon,\alpha}(f) \propto 1/f^\alpha$ with $\alpha = 0.7$ \cite{Dial2013}. While the charge noise spectrum has not been measured above a few MHz in GaAs or Si, higher frequency regimes are still important to the gate dynamics. Therefore, we extrapolate the spectra with a cautiously optimistic scenario ($\alpha = 0.7$) and a pessimistic scenario ($\alpha = 0$, i.e. white noise). We match each extrapolated spectrum to $S_0 = \SI{4e-20}{V^2/Hz}$ at \SI{1}{MHz} \cite{Dial2013}. For quasistatic charge and hyperfine noise we use the parameters given in \reftab{tab:params}, assuming $\sigma_{\db{}} = 0$ as a best case for Si. In GaAs, typical experiments work with \db{ij} from \SI{0.1}{ns^{-1}} to \SI{7}{ns^{-1}} using DNP \cite{Nichol2017}. In Si, gate voltage tuning of the electron g-factor \cite{Veldhorst2014} or micromagnets \cite{Wu2014} do not always allow for large gradients. Thus, we use values between \SI{0.01}{ns^{-1}} and \SI{0.7}{ns^{-1}}. Leakage can be suppressed by ensuring $\db{23} \gg J_{23}$, making spin exchange across dots 2 and 3 energetically costly \cite{Wardrop2014}. For interqubit capacitive coupling we consider two extreme cases, $\ec = 0$ and $\ec = \SI{350}{\micro eV}$. The latter is estimated from a typical charge stability diagram as the distance of the triple points belonging to the $(1,0)-(2,1)$ transition \cite{Botzem2018}.

\begin{table}
        \centering
        \renewcommand\arraystretch{1.25}
        \begin{tabular}{lrrrllrrr}
        & GaAs & Si & $~$ & & GaAs & Si\\
        \cline{1-3}\cline{5-7}
        $\sigma_{\eps}  \, \mr{(\mu V)}$ & \textsuperscript{\cite{Dial2013}}$8$     & 8  &   & $\sigma_{\db{}} \, \mr{(mT)}$     & 0.3                              & 0         \\
        $\epsilon_0 \, \mr{(mV)}$        & \textsuperscript{\cite{Dial2013}}$0.272$ & 0.272 & & $J_0 \, \mr{(ns^{-1})}$           & 1                                & 0.1         \\
        $\eps_\mr{min}$                  & $-5.4\eps_0$              & $-5.4\eps_0$ & & $\eps_\mr{max}$                   & $2.4\eps_0$                      & $2.4\eps_0$ \\
        $\fs \, \mr{(GS/s)}$             & 1                         & 0.1    &      & $\bg  \, (\mr{mT})$               & \textsuperscript{\cite{Cerfontaine2016}}$500$   & 500 \\
        \cline{1-3}\cline{5-7}
        \end{tabular}

\caption{Experimental parameters for GaAs and Si. Units of inverse seconds denote angular frequencies. For slower gates in Si we scale the impulse response by a factor $10$ in time, leading to \fs and $J_0$ given here.}
\label{tab:params}
\end{table}

\section{Two-qubit gates}
We now use our optimization strategy to search for a CNOT gate. The LMA finds solutions with high probability for $\nseg \ge 30$, typically within $10^4$ iterations, given our objective function and constraints. We calculate the gate fidelities post-optimization using the noise profiles described above, and show a representative pulse sequence in \reffig{fig:main}(b), with $\nseg = 50$ and $\ec = 0$. The pulse duration is \SI{50}{ns} and the field gradients are $\db{12} = -\db{34} = \SI{1}{ns^{-1}}$ and $\db{23} = \SI{7}{ns^{-1}}$. A large \db{23} was chosen to suppress leakage. This pulse exhibits a fidelity of \SI{99.90}{\%} for $\alpha = 0.7$ and small leakage, $\leak = \SI{1.9e-5}{}$, while unitary errors are negligible by design of the objective function. We list the different infidelity contributions for various parameter sets in \reftab{tab:best_cnot_fid}. The first two columns show that for $\alpha = 0$ the fast charge noise (\ifef) contribution is dominant due to the higher noise level above \SI{1}{MHz}, while for $\alpha = 0.7$ slow charge noise (\ifes) is dominant. In both cases, the infidelity is not limited by hyperfine noise (\ifb). We assess the gate's performance for different noise strengths in \reffig{fig:best_cnot_noise_scaling_ec0} and find that the infidelity contributions scale quadratically over a wide range of $\sigma_{\mr{\db{i}}}$, $\sigma_{\eps_{ij}}$ and $\sqrt{S_{\mathrm{\epsilon}_{ij}, 0.7}}$. However, as we discuss further in \refsec{sec:C} a fourth order term becomes dominant for large noise strengths, indicating partial dynamical decoupling from slow noise. We speculate that the quadratic contributions could be reduced by improving the dynamical decoupling. For $\db{23} \gg J_{23}$, noise on \db{23} has a far lesser effect on the gate's performance than intraqubit gradient noise, indicating that one need not stabilize \db{23} in GaAs. When we repeat the CNOT optimization for $\ec = \SI{350}{\micro eV}$ we find similar results (\SI{99.92}{\%} for $\alpha = 0.7$), suggesting that undesired capacitive coupling can be compensated by appropriate pulses. Thus we only consider $\ec = 0$ for the further analysis of the GaAs CNOT gate.

To investigate the optimal gate speed, \reffig{fig:main}(c) shows the scaling of the noise contributions with the AWG's sample rate \fs for $\alpha = 0.7$ obtained without reoptimization. We only adjust the time and energy scales while keeping all noise strengths fixed. Thus, they serve as a lower bound for the achievable fidelities. In the presence of only hyperfine noise (yellow curve), faster gates are advantageous. The infidelity from hyperfine noise scales with $\fs^4$ for lower and $\fs^2$ for higher fidelities, consistent with the dependence on the noise strength. The infidelity from slow charge noise is indifferent to the gate speed, related to the fact that the number of coherent exchange oscillations is constant with respect to the detuning \cite{Dial2013}. For pink noise we find $\ifef \propto \fs^{1-\alpha}$, which is consistent with $T_2(\eps)\propto(dJ/d\eps)^{-2}$ for white noise. The blue curve combines all noise contributions for $\alpha = 0.7$. It indicates that even though the optimization was performed for $\alpha = 0$, the original sample rate $\fs = \SI{1}{GS/s}$ is nearly optimal. 

Isotopically purified Si devices may prefer slower gate speeds because the lower magnetic field noise would lower the yellow curve in \reffig{fig:main}(c). Since slower control allows for lower field gradients, we use $\db{12} = -\db{34} = \SI{0.1}{ns^{-1}}$ and $\db{23} = \SI{0.7}{ns^{-1}}$. The analysis of the optimized \SI{500}{ns} long gate is shown in the last two columns of \reftab{tab:best_cnot_fid} for $\ec = \SI{350}{\micro eV}$. Its higher resilience to quasistatic charge noise indicates more effective dynamical decoupling. The total infidelity is about a factor 10 lower than in GaAs, reaching \SI{99.99}{\%} for $\alpha = 0.7$. If $S_0$ stays fixed at \SI{1}{MHz}, we expect a further improvement for $\alpha = 1$, which was observed in Si for lower frequencies up to \SI{320}{kHz} \cite{Yoneda2017}.

\begin{table}
\centering
\renewcommand\arraystretch{1.25}
\begin{tabular}{l *{4}{S[table-number-alignment=center,table-text-alignment=left,table-format=+1.1e+3,round-mode=figures,round-precision=2]}}
& \multicolumn{2}{c}{GaAs ($T = \SI{50}{ns}$)} & \multicolumn{2}{c}{Si ($T = \SI{500}{ns}$)} \\
& \multicolumn{1}{l}{$\alpha = 0$} & \multicolumn{1}{l}{$\alpha = 0.7$} & \multicolumn{1}{l}{$\alpha = 0$} & \multicolumn{1}{l}{$\alpha = 0.7$} \\
\hline
\ifes & 4.91e-04 & 5.04e-04 & 4.71e-05 & 4.85e-05 \\
\ifef & 1.56e-03 & 6.31e-05 & 2.40e-04 & 4.92e-05 \\
\ifb & 1.72e-04 & 1.68e-04 & 0 & 0 \\
\infid & 2.40e-03 & 9.76e-04 & 3.07e-04 & 1.09e-04 \\
\li & 1.14e-04 & 1.32e-06 & 6.43e-05 & 2.87e-06 \\
\lc & 1.80e-05 & 1.80e-05 & 1.04e-05 & 1.04e-05 \\
\leak & 1.32e-04 & 1.94e-05 & 7.47e-05 & 1.33e-05 \\
\hline
\end{tabular}
\caption{Analysis of the CNOT gate for GaAs and Si parameters, with $\ec = 0$ and \SI{350}{\micro eV}, respectively. The table lists the infidelity and leakage contributions as defined in the text for two spectral noise densities $S_{\epsilon,\alpha}(f) \propto 1/f^\alpha$ with $S_{\epsilon,\alpha}(\SI{1}{MHz}) = \SI{4e-20}{V^2/Hz}$ (calculated using 1000 noise realizations).}
\label{tab:best_cnot_fid}
\end{table}

\section{Single-qubit gates}
For closely spaced singlet-triplet qubits, residual interqubit exchange and capacitive coupling can complicate the parallel execution of single qubit gates. Thus, we optimize generators of the single-qubit Clifford group, $\ut = X_{\pi/2}\otimes\eye$ and $\ut = Y_{\pi/2}\otimes\eye$ choosing $\nseg = 20$ (as Ref.\,\cite{Cerfontaine2014}). We find similar fidelities for both gates and thus only discuss the results for $X_{\pi/2}\otimes\eye$ here. The results for $Y_{\pi/2}\otimes\eye$ are presented in \refsec{sec:C}. Since the exponential model for the exchange interaction has not been validated at small detuning, we do not expect such a model to capture the effects of residual interqubit exchange well when $\eps_{23}$ is at its minimum. Indeed, our measurements at the qubit's minimum detuning discussed in \refsec{sec:A} find a much larger residual exchange than predicted by the exponential model, $\je{ij} \ge \SI{0.005}{ns^{-1}}$. 

For the optimization, we keep $J_{23} \le \SI{0.005}{ns^{-1}}$ constant, leading to negligible leakage $\leak \le \SI{3e-7}{}$. Without capacitive coupling and residual $J_{23}$, we find a similar fidelity for GaAs (Si) parameters as for the CNOT gate, specifically \SI{99.93}{\%} (\SI{99.99}{\%}) for $\alpha = 0.7$. Large capacitive coupling of $\ec = \SI{350}{\micro eV}$ adds a systematic error of $\sim \SI{0.4}{\%}$ to the infidelity. The error due to residual exchange scales quadratically with $J_{23}$, and reaches $\sim 10^{-3}\,\%$ for $J_{23} = \SI{0.005}{ns^{-1}}$. By including residual coupling in the model for the optimization, we recover a fidelity of \SI{99.69}{\%} (\SI{99.95}{\%} for Si) as presented in \reftab{tab:best_xid_fid}. The resulting $X_{\pi/2}\otimes\eye$ pulse is shown in \reffig{fig:main}(d). It exhibits a non-idling identity by pulsing $J_{34}$ to suppress noise via dynamical decoupling. The capacitive effects are compensated by keeping the product of $J_{12}$ and $J_{34}$ low since the exponential model for $J_{ij}$ implies $\frac{\partial J_{ij}}{\partial\eps_{ij}} \propto J_{ij}$ and thus $\hc \propto J_{12} J_{34}$. This explains the pulses' interleaving nature. 

\section{ Implications}
Our work shows that if exchange coupling is used to mediate two-qubit gates, interqubit Coulomb coupling must be considered but presents no major obstacle for high-fidelity control of spin-qubit arrays. Using parameters of current GaAs (Si) devices, single-qubit gate fidelities of at least \SI{99.69}{\%} (\SI{99.95}{\%}) and two-qubit gate fidelities of \SI{99.90}{\%} (\SI{99.99}{\%}) are attainable. We find scaling laws to extrapolate these fidelities to different gate durations and noise strengths, which is useful to assess different material systems and gains from improving noise levels. However, to obtain the best fidelity, detailed knowledge of the high-frequency charge noise spectrum is beneficial.

Our approach can be extended to assess larger structures, scalable control hardware, and to determine the implication of qubit inhomogeneities on gate fidelities. Furthermore, it should be possible to implement our pulses experimentally using in-situ tune-up procedures~\cite{Cerfontaine2019a} to remove errors due to systematic inaccuracies in the model. Thus, our results are a strong indication that even in GaAs, which is attractive for optical coupling \cite{Joecker2018}, \sts qubits can provide a complete high-fidelity gate set for universal quantum computation.\textsl{}

\section{Acknowledgments}
We acknowledge helpful discussions with Tobias Hölzer. This work was supported by computing resources granted by RWTH Aachen University under project rwth0279, the European Research Council (ERC) under the European Union’s Horizon 2020 research and innovation program (grant agreement No. 679342), the Impulse and Networking Fund of the Helmholtz Association and Deutsche Forschungsgemeinschaft under Grant No. BL 1197/4-1. P.C. acknowledges support by Deutsche Telekom Stiftung. M.A.W. acknowledges the support of the U.S. Fulbright student program.

Reprinted with permission from Pascal Cerfontaine, René Otten, M. A. Wolfe, Patrick Bethke, and Hendrik Bluhm, Physical Review B 101, 155311 (2020). Copyright 2020 by the American Physical Society.

\appendix

\section{Model}
\label{sec:A}
\subsection{Pulse shaping}
\noindent In any experimental setup, qubit  control pulses are affected by finite rise times, ringing of the control hardware, and dispersion in the cables connecting the qubit. To account for these pulse distortions, we convolute ideal piece-wise constant pulses $\eps_{ij, k}$ with the measured impulse response of a typical experimental setup. 

Specifically, we used a Tektronix 5014C AWG in the amplified output mode to generate a square pulse with a rise time of $\approx \SI{1}{ns}$. This pulse is sent through \SI{150}{cm} of low-loss coaxial cable at room temperature, and a cryogenic coaxial cable assembly consisting of \SI{95}{cm} UT85 SS-BeCu cable, \SI{47}{cm} UT85 NbTi-NbTi cable, and \SI{28}{cm} UT85 Cu cable. The cryogenic wiring also includes three attenuator stages with a total of \SI{33}{dB} specified attenuation. This yields the measured square pulse shown in \reffig{fig:step_response} whose time-derivative is then used as a convolution kernel for $\eps_{ij, k}$ at each iteration of the optimization.

\begin{figure}[b]
	\includegraphics[width=\columnwidth]{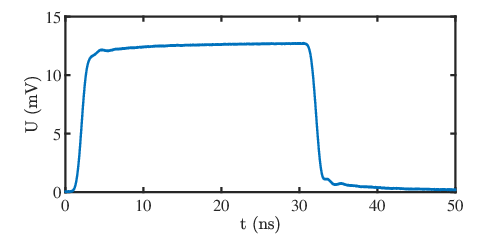}
	\caption{Distorted square pulse measured by sending a \SI{600}{mV} peak-to-peak square pulse across high-frequency coaxial cables and \SI{33}{dB} attenuation to the qubit sample holder at room temperature.}
	\label{fig:step_response}
\end{figure}

\subsection{Residual exchange interaction}
\noindent In order to extract the residual exchange $J_{\min}$ in the low-detuning regime, we employ the following measurement scheme on a GaAs double quantum dot (DQD) device identical to the second sample used by \citet{Cerfontaine2019ex}. Our method is based on extracting the oscillation frequencies of multiple free induction decay (FID) curves for a range of randomly fluctuating \dbz values. By observing oscillations where \dbz is close to zero, we can give an upper bound on $J_{\min}$ since the oscillation frequency is $\sqrt{\dbz^2 + J_{\min}^2}$.

We first initialize a singlet and quickly pulse deep into the (1,1) charge state for a variable time $\tau$, then read out the final state in the $\mr{S}$-$\mr{T}_0$ basis. Deep in (1,1) the residual exchange $J_{\min}$ is very small compared to most of the random values of \dbz but still greater than zero. After repeating the measurement for different evolution times $\tau$, we can thus extract the oscillation frequency $\omega = \sqrt{J_{\min}^2 + \dbz^2}$ by fitting $A\cos(\omega \tau + \phi)$.

Since these frequencies are almost always dominated by \dbz, we increase the chances of extracting an FID curve with low \dbz by using standard dynamic nuclear polarization based on appropriately calibrated $\mr{S}$-$\mr{T}_+$ Landau-Zener sweeps \cite{Bluhm2010}. We switch between $\mr{S}\rightarrow \mr{T}_+$ and $\mr{T}_+\rightarrow \mr{S}$ pumping (which either increase or decrease the average \dbz across the DQD) in order to tune the \dbz field through a zero-crossing. 
The lowest frequency we observe is $\omega = \SI{0.009}{ns^{-1}}$. Since the amplitude of this oscillation is still more than half of the maximum amplitude observed for higher \dbz, we can infer that $\dbz$ is still larger than $J_{\min}$. Thus, we extract $\SI{0.005}{ns^{-1}}$ as an upper bound for $J_{\min}$.

\subsection{Capacitive coupling}
\noindent We use a phenomenological model for the capacitive coupling Hamiltonian as given by Eq.\,2 since this allows us to include the experimentally observed exponential relation for \je{ij}. So far, no analytic model has been able to explain this exponential dependence over the full range of \je{ij}. In addition to being experimentally verified, using the phenomenological model allows cross checks with the experiment by \citet{Shulman2012}. The phenomenological model becomes invalid when $J_{23}$ is comparable to $J_{12}$ and $J_{34}$. When this is the case, interqubit tunneling becomes large and other charge configurations not present in the experiment of \citet{Shulman2012} alter the electrostatic coupling. 

For the single-qubit gates, $J_{23}$ is always at its minimum value, however it becomes large for short time intervals of the CNOT gate. While the phenomenological model may not be very accurate for the CNOT gate, we expect that it is adequate for assessing the size of systematic errors due to capacitive coupling, which we find to be on the order of \SI{0.4}{\%} (see \refsec{sec:C}). Since in-situ calibration routines \cite{Cerfontaine2014, Cerfontaine2019a, Cerfontaine2019ex} have shown to remove single-qubit gate errors as large as \SI{10}{\%}, we expect that errors in the capacitive coupling model can be similarly suppressed, even if the model does not capture the effect with a high accuracy.

\section{Gate optimization}
\label{sec:B}
\subsection{Convergence towards global minimum}
\noindent Since a single iteration of the Levenberg-Marquardt algorithm is not guaranteed to find a global minimum, we start the optimization from many different seeds. To assess to what extent this search is exhaustive, we bin the final optimized infidelities found for different initial random seeds for the CNOT gate presented in the main text (using GaAs parameters, $\ec = 0$ and $\alpha = 0$). The resulting histogram is shown in \reffig{fig:hist}. The sharp reduction in counts for gate fidelities above \SI{99.7}{\%} suggests that only small improvements can be expected when increasing the number of seeds, which should eventually lead to an exhaustive search.

We have also searched for shorter gates with $\nseg = 30$ segments and a gate time of $T = \SI{30}{ns}$. As expected the reduction in free parameters leads to faster convergence but lower fidelities of \SI{99.6}{\%}.

\begin{figure}[t!]
	\includegraphics[]{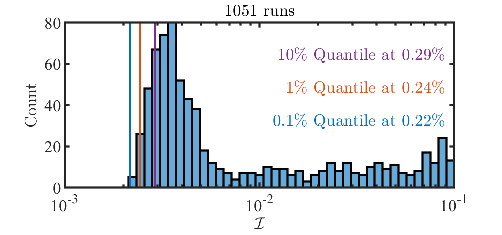}    
	\caption{Histogram of the final optimized infidelities found for different initial random seeds for the CNOT gate presented in the main text (using GaAs parameters, $\ec = 0$ and $\alpha = 0$). The colored lines represent \SI{0.1}{\%}, \SI{1}{\%} and \SI{10}{\%} quantiles. The infidelity shown here includes all infidelities due to noise but no leakage.}
	\label{fig:hist}
\end{figure}

\subsection{Markov approximation}
\noindent In the gate optimization we use a master equation approach combined with a Markov approximation to compute the infidelity due to fast uncorrelated charge noise in a computationally efficient manner.

Following \citet{Havel2003}, we start with the following Lindblad equation for a time-dependent density matrix $\rho = \rho(t)$ with the Lindbladian \lind, the time-independent Lindblad operators $L_k$, and Hamiltonian $H$:
\begin{align*}
\dot{\rho(t)} =& \lind(\rho) \\
=& i[\rho, H] + \sum_{k=0}^{K} \left( L_k \rho L_k^\dagger - \frac{1}{2} L_k^\dagger L_k \rho - \frac{1}{2} \rho L_k^\dagger L_k \right).
\end{align*}
This equation can also be written using superoperators acting on a vectorized density matrix $\vec{\rho}(t)$:
\begin{align*}
\vec{\rho}(t)    &= \exp\left[(-i\mathcal{H}+\mathcal{G})t\right] \vec{\rho}(0), \\[12pt]
\mathcal{H}      &= \eye \otimes H - H^\star \otimes \eye, \\
\mathcal{G}      & = \sum_{k=0}^{K} \mathcal{D}(L_k), \\
\mathcal{D}(L) &= L^\star \otimes L  -  \frac{1}{2} \eye \otimes(L^\dagger L)  -  \frac{1}{2}(L^{\star \dagger} L^\star) \otimes \eye.
\end{align*}
We use a Markov approximation and by summing over nearest neighbors we obtain the Lindblad equation given by \citet{Wardrop2014}
\begin{align*}
\mathcal{G} = \sum_{\langle i,j \rangle} S_{ij} \biggr\rvert\frac{dJ_{ij}}{d\eps_{ij}}\biggr\rvert^2 \mathcal{D}(\vecsig_i \cdot \vecsig_j), 
\end{align*}
where $S_{ij}$, defined by $\langle \eps_{ij}(t) \eps_{ij}(t+\Delta t)\rangle = S_{ij} \delta(\Delta t)$, is the charge noise spectral density and $\frac{dJ_{ij}}{d\eps_{ij}}$ is the sensitivity of the exchange interaction to charge noise. One can then use standard formulas \cite{Nielsen2002} to calculate the infidelity due to $\exp\left[(-i\mathcal{H}+\mathcal{G})t\right]$.

\subsection{Analytic derivatives}
\noindent To speed up the optimization significantly, we use analytic derivatives of $\uc$ and \qpef to efficiently differentiate all terms in the minimization problem from the main text with respect to $\eps_{ij,k}$. In this section we first describe how we calculate derivatives of unitary operators. This approach can also be generalized to the master equation used for calculating the effect of high-frequency charge noise.

We take an approach similar to \citet{Khaneja2005} and consider a time-independent $H_c$ with evolution time $\tau$. However, instead of their approximation for small $\tau$ in Eq.\,11 of Ref. \cite{Khaneja2005}, we base our derivative calculation on a first-order Magnus expansion of the unitary operator followed by analytic integration. We start by adding a small time-independent perturbation $H_v(\alpha)$ as a function of some time-independent parameter $\alpha$ to a time-independent Hamiltonian
\begin{align*}
H = H_c + H_v(\alpha).
\end{align*}
The total unitary after a time $t$ can be written as
\begin{align*}
U(t, \alpha) = U_c(t) \tilde{U}_v(t, \alpha),
\end{align*}
where $U_c(t)$ describes the time evolution due to $H_c$. $\tilde{U}_v(t, \alpha)$ solves the Schrödinger equation
\begin{align*}
i \frac{d\tilde{U}_v(t, \alpha)}{dt} = \tilde{H}_v (t, \alpha) \tilde{U}_v(t, \alpha),
\end{align*}
with $\hbar = 1$ and the interaction picture Hamiltonian
\begin{align*}
\tilde{H}_v(t, \alpha) = U_c^\dagger(t) H_v(\alpha) U_c(t).
\end{align*} 
Since $H_v(\alpha)$ is just an infinitesimally small perturbation with respect to $H_c$, we express $\tilde{U}_v(t)$ using first order Magnus expansion. If $H_v(\alpha) = \alpha M$, we can calculate $d U(\tau, \alpha) / d \alpha$ using the chain rule.

To extend this approach to a time-dependent $H_c(t)$ we follow \citet{Khaneja2005} and approximate $H_c(t)$ as piecewise constant with $n$ segments. We can then calculate the derivative with respect to $\alpha^{(k)}$ for each discrete time step $k$. To obtain the derivative of the full unitary we utilize the fact that the total control unitary is a product of shorter control gates at time step $k$,
\begin{align*}
U_c(\tau) =  \prod_{j=n}^1 U_c^{(k)}.
\end{align*}
We then obtain the derivative of the full unitary with respect to $\alpha^{(k)}$,
\begin{align*}
\frac{d U_c(\tau)}{d \alpha^{(k)}} =  U_{\mr{future}}^{(k)}\frac{\partial U_c^{(k)}}{\partial \alpha^{(k)}}U_{\mr{past}}^{(k)},
\end{align*}
where $U_{\mr{past}}^{(k)} = \prod_{j=k-1}^1 U_c^{(j)}$ and $U_{\mr{future}}^{(k)} = \prod_{j=n}^{k+1} U_c^{(j)}$. 

When using the Hamiltonian from Eq.\,1 as the control Hamiltonian $H_c$, the derivatives are computed by replacing $\alpha^{(k)}$ by $J_{ij,k}$. This also requires a different $M$ for each $J_{ij,k}$, which can be directly extracted from Eq.\,1. For $J_{ij,k}$ this directly yields the derivatives $d U_c(\tau)/ d J_{ij,k}$. In order to obtain analytic derivatives of the components of the minimization problem from the main text, we employ the chain rule.

\section{Gate analysis}
\label{sec:C}
\noindent In this section, we show additional data and analysis on the single- and two-qubit gates, calculated using 1000 Monte Carlo time traces with $\approx \SI{3}{\%}$ relative error. Due to statistical fluctuations and interaction effects, the separately calculated total infidelity $\mathcal{I}$ is not exactly equal to the sum of the different infidelity contributions. To estimate the infidelity from fast charge noise, we always use frequency cutoffs at $1/T$ and \SI{100}{GHz}.

\begin{table}
	\centering
	\renewcommand\arraystretch{1.25}
	\begin{tabular}{l *{4}{S[table-number-alignment=center,table-text-alignment=left,table-format=+1.1e+3,round-mode=figures,round-precision=2]}}
		& \multicolumn{2}{c}{GaAs ($T = \SI{20}{ns}$)} & \multicolumn{2}{c}{Si ($T = \SI{200}{ns}$)} \\
		& \multicolumn{1}{l}{$\alpha = 0$} & \multicolumn{1}{l}{$\alpha = 0.7$} & \multicolumn{1}{l}{$\alpha = 0$} & \multicolumn{1}{l}{$\alpha = 0.7$} \\
		\hline
		\ifes & 1.46e-03 & 1.46e-03 & 4.44e-04 & 4.62e-04 \\
		\ifef & 1.96e-03 & 5.97e-05 & 4.20e-04 & 5.96e-05 \\
		\ifb & 1.50e-03 & 1.52e-03 & 0 & 0 \\
		\infid & 5.06e-03 & 3.05e-03 & 8.52e-04 & 5.18e-04 \\
		\li & 2.32e-08 & 2.01e-09 & 3.19e-08 & 3.24e-09 \\
		\lc & 1.09e-07 & 1.09e-07 & 2.67e-07 & 2.67e-07 \\
		\leak & 1.32e-07 & 1.11e-07 & 2.99e-07 & 2.70e-07 \\
		\hline
	\end{tabular}
	\caption{Analysis of the residual coupling compensating $X_{\pi/2}\otimes\eye$ gates for $J_{23} = \SI{0.005}{ns^{-1}}$ and large capacitive coupling $\ec = \SI{350}{\micro eV}$. Other parameters are the same as in Tab.\,II in the main text. The total fidelities calculated with all noise sources applied simultaneously are \SI{99.49}{\%}, \SI{99.69}{\%}, \SI{99.91}{\%} and \SI{99.95}{\%} (from left to right). Leakage is much smaller than for the CNOT gate since $J_{23} = \SI{0.005}{ns^{-1}}$ is almost turned off.}
	\label{tab:best_xid_fid}
\end{table}

\begin{figure}[t!]
	\includegraphics{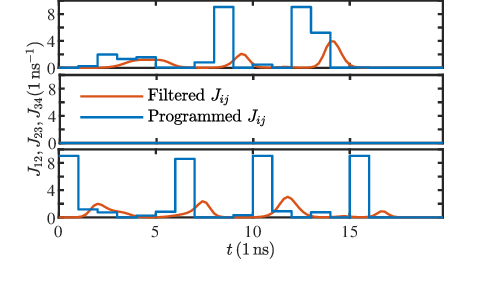}
	\caption{Residual coupling compensating $Y_{\pi/2}\otimes\eye$ pulse sequence optimized for $\ec = \SI{350}{\micro eV}$ and $J_{23} = \SI{0.005}{ns^{-1}}$ with \SI{99.82}{\%} fidelity in GaAs.}
	\label{fig:best_yid_pulse_ec350}
\end{figure}

\begin{table}[h!]
	\centering
	\renewcommand\arraystretch{1.25}
	\begin{tabular}{l *{4}{S[table-number-alignment=center,table-text-alignment=left,table-format=+1.1e+3,round-mode=figures,round-precision=2]}}
		& \multicolumn{2}{c}{GaAs ($J_{23} = 0, \ec = 0$)} & \multicolumn{2}{c}{GaAs ($J_{23} > 0, \ec > 0$)} \\
		& \multicolumn{1}{l}{$\alpha = 0$} & \multicolumn{1}{l}{$\alpha = 0.7$} & \multicolumn{1}{l}{$\alpha = 0$} & \multicolumn{1}{l}{$\alpha = 0.7$} \\
		\hline
		\ifes & 3.48e-04 & 3.49e-04 & 9.59e-04 & 9.70e-04 \\
		\ifef & 1.69e-03 & 5.62e-05 & 2.04e-03 & 6.17e-05 \\
		\ifb & 1.40e-04 & 1.39e-04 & 6.78e-04 & 7.09e-04 \\
		\infid & 2.17e-03 & 5.66e-04 & 3.78e-03 & 1.79e-03 \\
		\li & 0 & 0 & 3.39e-08 & 1.28e-09 \\
		\lc & 0 & 0 & 1.84e-07 & 1.84e-07 \\
		\leak & 0 & 0 & 2.18e-07 & 1.85e-07 \\
		\hline
	\end{tabular}
	\caption{Analysis of a $Y_{\pi/2}\otimes\eye$ gate. For the first two columns, $J_{23} = 0$ and $\ec = 0$ were used in the optimization, while $J_{23} = \SI{0.005}{ns^{-1}}$ and $\ec = \SI{350}{\micro eV}$ were used in the last two columns. Other parameters are the same as in the main text. The total fidelities calculated with all noise sources applied simultaneously are \SI{99.78}{\%}, \SI{99.94}{\%}, \SI{99.62}{\%} and \SI{99.82}{\%} (from left to right).}
	\label{tab:best_yid}
\end{table}

\begin{table}[h!]
	\centering
	\renewcommand\arraystretch{1.25}
	\begin{tabular}{l *{4}{S[table-number-alignment=center,table-text-alignment=left,table-format=+1.1e+3,round-mode=figures,round-precision=2]}}
		& \multicolumn{2}{c}{GaAs ($T = \SI{20}{ns}$)} & \multicolumn{2}{c}{Si ($T = \SI{200}{ns}$)} \\
		& \multicolumn{1}{l}{$\alpha = 0$} & \multicolumn{1}{l}{$\alpha = 0.7$} & \multicolumn{1}{l}{$\alpha = 0$} & \multicolumn{1}{l}{$\alpha = 0.7$} \\
		\hline
		\ifes & 5.01e-04 & 5.08e-04 & 4.63e-05 & 6.03e-05 \\
		\ifef & 1.89e-03 & 5.74e-05 & 2.58e-04 & 4.60e-05 \\
		\ifb & 1.18e-04 & 1.24e-04 & 0 & 0 \\
		\infid & 2.59e-03 & 7.14e-04 & 3.11e-04 & 1.06e-04 \\
		\li & 0 & 0 & 0 & 0 \\
		\lc & 0 & 0& 0 & 0 \\
		\leak & 0 & 0 & 0 & 0 \\
		\hline
	\end{tabular}
	\caption{Analysis of a $X_{\pi/2}\otimes\eye$ gate for $J_{23} = 0$ and no capacitive coupling $\ec = 0$. Other parameters are the same as in the main text. The total fidelities calculated with all noise sources applied simultaneously are \SI{99.74}{\%}, \SI{99.93}{\%}, \SI{99.97}{\%} and \SI{99.99}{\%} (from left to right).
	}
	\label{tab:best_xid_ec0}
\end{table}

\begin{figure}[h!]
	\includegraphics{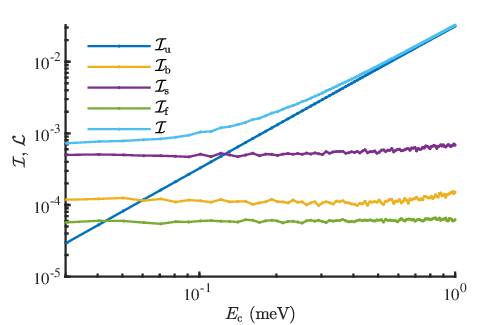}
	\caption{Impact of \ec on the $X_{\pi/2}\otimes\eye$ pulse from \reftab{tab:best_xid_ec0} using GaAs parameters and $\alpha = 0.7$. The pulse was optimized for $\ec = 0$ and $J_{23} = 0$. Larger values of \ec mostly cause systematic errors $\ifu = 1 - \fid(\uc, \ut)$, the infidelities from noise increase only towards very large values on the order of \SI{1}{meV}. For $\ec = \SI{350}{ \mu eV}$ the unitary infidelity \ifu is about \SI{0.4}{\%}.}
	\label{fig:best_xid_pulse_ec0_ec_scaling}
\end{figure}

\begin{figure}[h!]
	\includegraphics{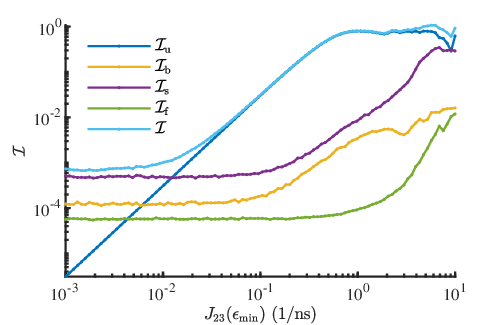}
	\caption{Impact of residual $J_{23}$ on the $X_{\pi/2}\otimes\eye$ pulse from \reftab{tab:best_xid_ec0} using GaAs parameters and $\alpha = 0.7$. The pulse was optimized for $\ec = 0$ and $J_{23} = 0$. As for \ec, larger values of $J_{23}$ mostly cause systematic errors, the infidelities from noise increase only towards very large values on the order of \SI{0.1}{ns^{-1}}. For $J_{23} = \SI{0.005}{ns^{-1}}$ the unitary infidelity \ifu is on the order of $10^{-3}\,\%$.}
	\label{fig:best_xid_pulse_ec0_j23_scaling}
\end{figure}

\subsection{Single-qubit gates}

\noindent We show a complete analysis of the fidelity and leakage contributions of the $X_{\pi/2}\otimes\eye$ gate from the main text in \reftab{tab:best_xid_fid}.  For completeness, we also present data on the $Y_{\pi/2}\otimes\eye$ gate. Its pulse shape is shown in \reffig{fig:best_yid_pulse_ec350}. The fidelities of the residual coupling compensating $Y_{\pi/2}\otimes\eye$ gate are listed in the last two columns of \reftab{tab:best_yid}, and are a little bit better than those of the $X_{\pi/2}\otimes\eye$ gate.

We also give the fidelity contributions for $X_{\pi/2}\otimes\eye$ and $Y_{\pi/2}\otimes\eye$ gates optimized with no residual coupling ($J_{23} = 0$ and $\ec = 0$) in \reftab{tab:best_xid_ec0} and the first two columns of \reftab{tab:best_yid}, respectively. As mentioned in the main text these are slightly better, indicating that higher fidelities may be attainable if parallel single-qubit gates are optimized in the same way as two-qubit gates (by allowing $J_{23}$ to be pulsed to higher values in the optimization).

To assess the effect of nonzero \ec and $J_{23}$ when they are not taken into account in the optimization, \reffig{fig:best_xid_pulse_ec0_ec_scaling} and \reffig{fig:best_xid_pulse_ec0_j23_scaling} show sweeps of \ec and $J_{23}$. For both sweeps we find that moderate values of \ec and $J_{23}$ predominantly cause systematic errors. The infidelities from noise increase only towards very large values when noise either couples in directly in Eq.\,2 or via a larger $\frac{\partial J_{23}}{\partial\eps_{23}}$ in Eq.\,1. For $\ec = \SI{350}{ \mu eV}$ the unitary infidelity $\ifu = 1 - \fid(\uc, \ut)$  is about \SI{0.4}{\%}. The effect of residual exchange is much lower, about $10^{-3}\,\%$ for $J_{23} = \SI{0.005}{ns^{-1}}$.

\subsection{Two-qubit gate}
\noindent For completeness, we list the CNOT infidelity contributions for $\ec = \SI{350}{\micro eV}$ in \reftab{tab:best_cnot_ec350uV}. These figures deviate only slightly from those presented in the main text, indicating that the optimization can fully account for different values of \ec. Furthermore, we give the pulse for the gate optimized with $\ec = \SI{350}{ueV}$ and GaAs parameters in \reffig{fig:best_cnot_pulse_ec350}, and for the Si gate in \reffig{fig:best_cnot_pulse_si_ec350}. We now present some further analysis, using the CNOT gate presented in the main text (with GaAs parameters and $\ec = 0$) as an example. 

\begin{table}[t!]
	\centering
	\renewcommand\arraystretch{1.25}
	\begin{tabular}{l *{4}{S[table-number-alignment=center,table-text-alignment=left,table-format=+1.1e+3,round-mode=figures,round-precision=2]}}
		& \multicolumn{2}{c}{GaAs ($T = \SI{50}{ns}$)} & \multicolumn{2}{c}{Si ($T = \SI{500}{ns}$)} \\
		& \multicolumn{1}{l}{$\alpha = 0$} & \multicolumn{1}{l}{$\alpha = 0.7$} & \multicolumn{1}{l}{$\alpha = 0$} & \multicolumn{1}{l}{$\alpha = 0.7$} \\
		\hline
		\ifes & 4.84e-04 & 4.58e-04 & 4.71e-05 & 4.85e-05 \\
		\ifef & 1.50e-03 & 6.45e-05 & 2.40e-04 & 4.92e-05 \\
		\ifb & 1.73e-04 & 1.59e-04 & 0 & 0 \\
		\infid & 2.30e-03 & 8.19e-04 & 3.07e-04 & 1.09e-04 \\
		\li & 6.41e-05 & 7.03e-07 & 6.43e-05 & 2.87e-06 \\
		\lc & 2.69e-05 & 2.69e-05 & 1.04e-05 & 1.04e-05 \\
		\leak & 9.10e-05 & 2.76e-05 & 7.47e-05 & 1.33e-05 \\
		\hline
	\end{tabular}
	\caption{Analysis of the CNOT gate for GaAs and Si parameters, with $\ec = \SI{350}{\micro eV}$. The total fidelities calculated with all noise sources applied simultaneously are \SI{99.76}{\%}, \SI{99.92}{\%}, \SI{99.96}{\%} and \SI{99.99}{\%} (from left to right), with small leakage and negligible systematic errors.}
	\label{tab:best_cnot_ec350uV}
\end{table}

\begin{figure}[t!]
	\includegraphics[width = 1\columnwidth,trim=0 10 0 0,clip]{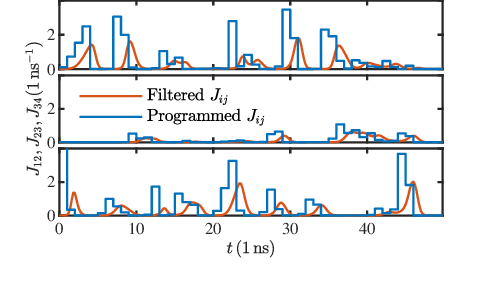}
	\caption{CNOT pulse sequence optimized for GaAs parameters and $\ec = \SI{350}{\micro eV}$ with \SI{99.92}{\%} fidelity in GaAs.}
	\label{fig:best_cnot_pulse_ec350}
	\vspace{1em}
	\includegraphics[width = 1\columnwidth,trim=0 10 0 0,clip]{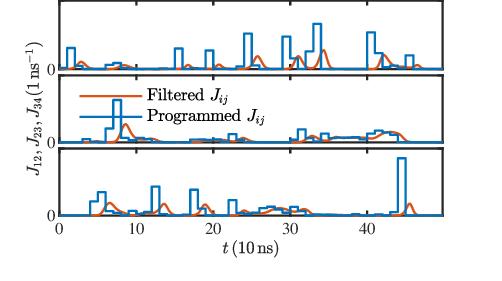}
	\caption{CNOT pulse sequence optimized for Si parameters and $\ec = \SI{350}{\micro eV}$ with \SI{99.99}{\%} fidelity in Si.}
	\label{fig:best_cnot_pulse_si_ec350}
\end{figure}

\begin{figure}[ht!]
	\includegraphics[width = 1\columnwidth]{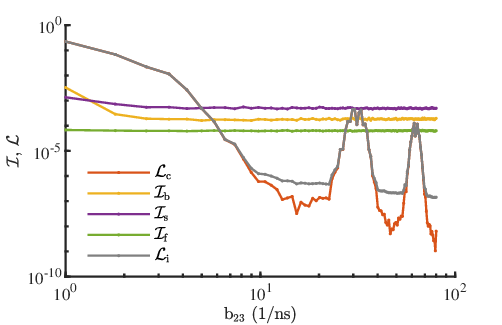}
	\caption{Leakage and infidelity contributions of the CNOT gate from the main text with varying \db{23}. This plot uses GaAs parameters, $\alpha = 0.7$, $\ec = 0$ and $\db{23} = \SI{7}{ns^{-1}}$. Note that $\lc > 0$ since the optimization algorithm balances all contributions to the objective function in the minimization problem.}
	\label{fig:best_cnot_b23_scaling_ec0}
\end{figure}

\begin{figure}[t!]
	\includegraphics[width = 1\columnwidth]{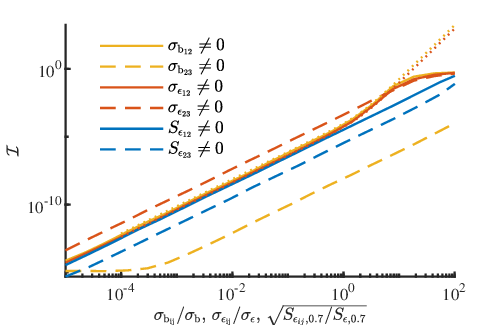}
	\caption{Analysis of the CNOT gate presented in the main text, using GaAs parameters, $\alpha = 0.7$ and $\ec = 0$. The varying noise amplitudes are given relative to the GaAs noise parameters. For each trace, all noise sources except the one indicated in the corresponding legend entry are turned off. The dotted yellow (red) line represents a fit to $\sigma_{\mathrm{b}_{12}} \neq 0$ ($\sigma_{\epsilon_{12}} \neq 0$) as explained in the text.}
	\label{fig:best_cnot_noise_scaling_ec0}
\end{figure}

First, we analyze the effect of the magnitude of \db{23} on leakage and the infidelity contributions as shown in \reffig{fig:best_cnot_b23_scaling_ec0}. As long as \db{23} is higher than \SI{2}{ns^{-1}}, the infidelity contributions do not show a significant dependence on \db{23}. Furthermore, both incoherent leakage \li and coherent leakage \lc can be suppressed by several orders of magnitude if $\db{23} \ge \SI{10}{ns^{-1}}$. The periodic increase in leakage can be attributed to the integration step of \SI{0.2}{ns} used in the numerical search. While we use a \SI{0.002}{ns} integration step in the final characterization, the pulses still contain \SI{0.2}{ns} steps in $J_{ij}$ from the original pulse shapes. While leakage is already low for the gates presented here, choosing $\db{23} = \SI{20}{ns^{-1}}$ would lead to a further reduction at the cost of a slightly higher experimental complexity.

In order to assess the effect for different samples and material systems, we also subject the gate to different noise strengths. The result is shown in \reffig{fig:best_cnot_noise_scaling_ec0} and indicates that noise on \db{23} has a four orders of magnitude lower effect than noise on \db{12} or \db{34} since it only enters the Hamiltonian in the leakage subspace. Furthermore, the infidelity contributions scale quadratically with $\sigma_{\mr{\db{i}}}$, $\sigma_{\eps_{ij}}$ and $\sqrt{S_{\mathrm{\epsilon}_{ij}, 0.7}}$ over a wide range. However, for relative noise strengths above $1$, we observe a fourth order scaling for quasistatic noise on $\eps_{12}$, $\eps_{34}$, \db{12} and \db{34}. This is typically observed for dynamically corrected gates (DCGs) which are first-order insensitive to slow noise. In order to investigate this further, we perform fits of the form $ax^2 + bx^4$, shown as a dotted yellow and red line in \reffig{fig:best_cnot_noise_scaling_ec0} for $\db{12}$ and $\eps_{12}$ noise, respectively. For $\db{12}$ noise, the fit coefficients are $a = \SI{7.5e-5}{}$ and $b= \SI{1.7e-5}{}$. For $\eps_{12}$ noise we obtain $a = \SI{4.3e-5}{}$ and $b = \SI{0.9e-5}{}$. These fits imply that our gates decouple partly from slow noise. This indicates that the quadratic scaling relation is not a fundamental limit but results from incomplete cancellation of slow noise. Once the corresponding infidelity contribution to the objective function (of the minimization problem from the main text) is substantially smaller than other terms, the algorithm does not reduce it any further, either because of incomplete convergence or because the cost in terms of other infidelity contributions would be larger than the very small benefit. A complete cancellation could be achieved by explicitly requiring dynamical decoupling in the numerical search, possibly at the cost of a slight increase of other contributions. A stronger cancellation may also explain why the value of $\ifes$ obtained by optimizing with Si parameters (purple triangle at $f_s=10^{-1}$ GS/s in Fig.\,2(c) of the main text) is substantially lower than the scaled result (purple line), which was not a dominant contribution in the optimization of the corresponding gate (see Tab.\,II, first column). Likewise, one might expect that further improvement is possible for the $\alpha = 0.7$ scenario in GaAs, which is dominated by $\ifes$, unlike the $\alpha =0$ case used to optimize the pulse.


\begin{thebibliography}{42}%
\makeatletter
\providecommand \@ifxundefined [1]{%
 \@ifx{#1\undefined}
}%
\providecommand \@ifnum [1]{%
 \ifnum #1\expandafter \@firstoftwo
 \else \expandafter \@secondoftwo
 \fi
}%
\providecommand \@ifx [1]{%
 \ifx #1\expandafter \@firstoftwo
 \else \expandafter \@secondoftwo
 \fi
}%
\providecommand \natexlab [1]{#1}%
\providecommand \enquote  [1]{``#1''}%
\providecommand \bibnamefont  [1]{#1}%
\providecommand \bibfnamefont [1]{#1}%
\providecommand \citenamefont [1]{#1}%
\providecommand \href@noop [0]{\@secondoftwo}%
\providecommand \href [0]{\begingroup \@sanitize@url \@href}%
\providecommand \@href[1]{\@@startlink{#1}\@@href}%
\providecommand \@@href[1]{\endgroup#1\@@endlink}%
\providecommand \@sanitize@url [0]{\catcode `\\12\catcode `\$12\catcode
  `\&12\catcode `\#12\catcode `\^12\catcode `\_12\catcode `\%12\relax}%
\providecommand \@@startlink[1]{}%
\providecommand \@@endlink[0]{}%
\providecommand \url  [0]{\begingroup\@sanitize@url \@url }%
\providecommand \@url [1]{\endgroup\@href {#1}{\urlprefix }}%
\providecommand \urlprefix  [0]{URL }%
\providecommand \Eprint [0]{\href }%
\providecommand \doibase [0]{http://dx.doi.org/}%
\providecommand \selectlanguage [0]{\@gobble}%
\providecommand \bibinfo  [0]{\@secondoftwo}%
\providecommand \bibfield  [0]{\@secondoftwo}%
\providecommand \translation [1]{[#1]}%
\providecommand \BibitemOpen [0]{}%
\providecommand \bibitemStop [0]{}%
\providecommand \bibitemNoStop [0]{.\EOS\space}%
\providecommand \EOS [0]{\spacefactor3000\relax}%
\providecommand \BibitemShut  [1]{\csname bibitem#1\endcsname}%
\let\auto@bib@innerbib\@empty
\bibitem [{\citenamefont {Yoneda}\ \emph {et~al.}(2017)\citenamefont {Yoneda},
  \citenamefont {Takeda}, \citenamefont {Otsuka}, \citenamefont {Nakajima},
  \citenamefont {Delbecq}, \citenamefont {Allison}, \citenamefont {Honda},
  \citenamefont {Kodera}, \citenamefont {Oda}, \citenamefont {Hoshi},
  \citenamefont {Usami}, \citenamefont {Itoh},\ and\ \citenamefont
  {Tarucha}}]{Yoneda2017}%
  \BibitemOpen
  \bibfield  {author} {\bibinfo {author} {\bibfnamefont {J.}~\bibnamefont
  {Yoneda}}, \bibinfo {author} {\bibfnamefont {K.}~\bibnamefont {Takeda}},
  \bibinfo {author} {\bibfnamefont {T.}~\bibnamefont {Otsuka}}, \bibinfo
  {author} {\bibfnamefont {T.}~\bibnamefont {Nakajima}}, \bibinfo {author}
  {\bibfnamefont {M.~R.}\ \bibnamefont {Delbecq}}, \bibinfo {author}
  {\bibfnamefont {G.}~\bibnamefont {Allison}}, \bibinfo {author} {\bibfnamefont
  {T.}~\bibnamefont {Honda}}, \bibinfo {author} {\bibfnamefont
  {T.}~\bibnamefont {Kodera}}, \bibinfo {author} {\bibfnamefont
  {S.}~\bibnamefont {Oda}}, \bibinfo {author} {\bibfnamefont {Y.}~\bibnamefont
  {Hoshi}}, \bibinfo {author} {\bibfnamefont {N.}~\bibnamefont {Usami}},
  \bibinfo {author} {\bibfnamefont {K.~M.}\ \bibnamefont {Itoh}}, \ and\
  \bibinfo {author} {\bibfnamefont {S.}~\bibnamefont {Tarucha}},\ }\href
  {\doibase 10.1038/s41565-017-0014-x} {\bibfield  {journal} {\bibinfo
  {journal} {Nat. Nanotechnol.}\ }\textbf {\bibinfo {volume} {13}},\ \bibinfo
  {pages} {1} (\bibinfo {year} {2017})}\BibitemShut {NoStop}%
\bibitem [{\citenamefont {Dehollain}\ \emph {et~al.}(2016)\citenamefont
  {Dehollain}, \citenamefont {Muhonen}, \citenamefont {Blume-Kohout},
  \citenamefont {Rudinger}, \citenamefont {Gamble}, \citenamefont {Nielsen},
  \citenamefont {Laucht}, \citenamefont {Simmons}, \citenamefont {Kalra},
  \citenamefont {Dzurak},\ and\ \citenamefont {Morello}}]{Dehollain2016}%
  \BibitemOpen
  \bibfield  {author} {\bibinfo {author} {\bibfnamefont {J.~P.}\ \bibnamefont
  {Dehollain}}, \bibinfo {author} {\bibfnamefont {J.~T.}\ \bibnamefont
  {Muhonen}}, \bibinfo {author} {\bibfnamefont {R.}~\bibnamefont
  {Blume-Kohout}}, \bibinfo {author} {\bibfnamefont {K.~M.}\ \bibnamefont
  {Rudinger}}, \bibinfo {author} {\bibfnamefont {J.~K.}\ \bibnamefont
  {Gamble}}, \bibinfo {author} {\bibfnamefont {E.}~\bibnamefont {Nielsen}},
  \bibinfo {author} {\bibfnamefont {A.}~\bibnamefont {Laucht}}, \bibinfo
  {author} {\bibfnamefont {S.}~\bibnamefont {Simmons}}, \bibinfo {author}
  {\bibfnamefont {R.}~\bibnamefont {Kalra}}, \bibinfo {author} {\bibfnamefont
  {A.~S.}\ \bibnamefont {Dzurak}}, \ and\ \bibinfo {author} {\bibfnamefont
  {A.}~\bibnamefont {Morello}},\ }\href {\doibase
  10.1088/1367-2630/18/10/103018} {\bibfield  {journal} {\bibinfo  {journal}
  {New J. Phys.}\ }\textbf {\bibinfo {volume} {18}},\ \bibinfo {pages} {103018}
  (\bibinfo {year} {2016})}\BibitemShut {NoStop}%
\bibitem [{\citenamefont {Veldhorst}\ \emph {et~al.}(2015)\citenamefont
  {Veldhorst}, \citenamefont {Yang}, \citenamefont {Hwang}, \citenamefont
  {Huang}, \citenamefont {Dehollain}, \citenamefont {Muhonen}, \citenamefont
  {Simmons}, \citenamefont {Laucht}, \citenamefont {Hudson}, \citenamefont
  {Itoh}, \citenamefont {Morello},\ and\ \citenamefont
  {Dzurak}}]{Veldhorst2015}%
  \BibitemOpen
  \bibfield  {author} {\bibinfo {author} {\bibfnamefont {M.}~\bibnamefont
  {Veldhorst}}, \bibinfo {author} {\bibfnamefont {C.~H.}\ \bibnamefont {Yang}},
  \bibinfo {author} {\bibfnamefont {J.~C.~C.}\ \bibnamefont {Hwang}}, \bibinfo
  {author} {\bibfnamefont {W.}~\bibnamefont {Huang}}, \bibinfo {author}
  {\bibfnamefont {J.~P.}\ \bibnamefont {Dehollain}}, \bibinfo {author}
  {\bibfnamefont {J.~T.}\ \bibnamefont {Muhonen}}, \bibinfo {author}
  {\bibfnamefont {S.}~\bibnamefont {Simmons}}, \bibinfo {author} {\bibfnamefont
  {A.}~\bibnamefont {Laucht}}, \bibinfo {author} {\bibfnamefont {F.~E.}\
  \bibnamefont {Hudson}}, \bibinfo {author} {\bibfnamefont {K.~M.}\
  \bibnamefont {Itoh}}, \bibinfo {author} {\bibfnamefont {A.}~\bibnamefont
  {Morello}}, \ and\ \bibinfo {author} {\bibfnamefont {A.~S.}\ \bibnamefont
  {Dzurak}},\ }\href {\doibase 10.1038/nature15263} {\bibfield  {journal}
  {\bibinfo  {journal} {Nature}\ }\textbf {\bibinfo {volume} {526}},\ \bibinfo
  {pages} {410} (\bibinfo {year} {2015})}\BibitemShut {NoStop}%
\bibitem [{\citenamefont {Watson}\ \emph {et~al.}()\citenamefont {Watson},
  \citenamefont {Philips}, \citenamefont {Kawakami}, \citenamefont {Ward},
  \citenamefont {Scarlino}, \citenamefont {Veldhorst}, \citenamefont {Savage},
  \citenamefont {Lagally}, \citenamefont {Friesen}, \citenamefont
  {Coppersmith}, \citenamefont {Eriksson},\ and\ \citenamefont
  {Vandersypen}}]{Watson2017}%
  \BibitemOpen
  \bibfield  {author} {\bibinfo {author} {\bibfnamefont {T.~F.}\ \bibnamefont
  {Watson}}, \bibinfo {author} {\bibfnamefont {S.~G.~J.}\ \bibnamefont
  {Philips}}, \bibinfo {author} {\bibfnamefont {E.}~\bibnamefont {Kawakami}},
  \bibinfo {author} {\bibfnamefont {D.~R.}\ \bibnamefont {Ward}}, \bibinfo
  {author} {\bibfnamefont {P.}~\bibnamefont {Scarlino}}, \bibinfo {author}
  {\bibfnamefont {M.}~\bibnamefont {Veldhorst}}, \bibinfo {author}
  {\bibfnamefont {D.~E.}\ \bibnamefont {Savage}}, \bibinfo {author}
  {\bibfnamefont {M.~G.}\ \bibnamefont {Lagally}}, \bibinfo {author}
  {\bibfnamefont {M.}~\bibnamefont {Friesen}}, \bibinfo {author} {\bibfnamefont
  {S.~N.}\ \bibnamefont {Coppersmith}}, \bibinfo {author} {\bibfnamefont
  {M.~A.}\ \bibnamefont {Eriksson}}, \ and\ \bibinfo {author} {\bibfnamefont
  {L.~M.~K.}\ \bibnamefont {Vandersypen}},\ }\href
  {http://arxiv.org/abs/1708.04214} {\ }\Eprint
  {http://arxiv.org/abs/1708.04214} {arXiv:1708.04214} \BibitemShut {NoStop}%
\bibitem [{\citenamefont {Zajac}\ \emph {et~al.}(2018)\citenamefont {Zajac},
  \citenamefont {Sigillito}, \citenamefont {Russ}, \citenamefont {Borjans},
  \citenamefont {Taylor}, \citenamefont {Burkard},\ and\ \citenamefont
  {Petta}}]{Zajac2017}%
  \BibitemOpen
  \bibfield  {author} {\bibinfo {author} {\bibfnamefont {D.~M.}\ \bibnamefont
  {Zajac}}, \bibinfo {author} {\bibfnamefont {A.~J.}\ \bibnamefont
  {Sigillito}}, \bibinfo {author} {\bibfnamefont {M.}~\bibnamefont {Russ}},
  \bibinfo {author} {\bibfnamefont {F.}~\bibnamefont {Borjans}}, \bibinfo
  {author} {\bibfnamefont {J.~M.}\ \bibnamefont {Taylor}}, \bibinfo {author}
  {\bibfnamefont {G.}~\bibnamefont {Burkard}}, \ and\ \bibinfo {author}
  {\bibfnamefont {J.~R.}\ \bibnamefont {Petta}},\ }\href {\doibase
  10.1126/science.aao5965} {\bibfield  {journal} {\bibinfo  {journal}
  {Science}\ }\textbf {\bibinfo {volume} {359}},\ \bibinfo {pages} {439}
  (\bibinfo {year} {2018})}\BibitemShut {NoStop}%
\bibitem [{\citenamefont {Huang}\ \emph {et~al.}()\citenamefont {Huang},
  \citenamefont {Yang}, \citenamefont {Chan}, \citenamefont {Tanttu},
  \citenamefont {Hensen}, \citenamefont {Leon}, \citenamefont {Fogarty},
  \citenamefont {Hwang}, \citenamefont {Hudson}, \citenamefont {Itoh},
  \citenamefont {Morello}, \citenamefont {Laucht},\ and\ \citenamefont
  {Dzurak}}]{Huang2018a}%
  \BibitemOpen
  \bibfield  {author} {\bibinfo {author} {\bibfnamefont {W.}~\bibnamefont
  {Huang}}, \bibinfo {author} {\bibfnamefont {C.~H.}\ \bibnamefont {Yang}},
  \bibinfo {author} {\bibfnamefont {K.~W.}\ \bibnamefont {Chan}}, \bibinfo
  {author} {\bibfnamefont {T.}~\bibnamefont {Tanttu}}, \bibinfo {author}
  {\bibfnamefont {B.}~\bibnamefont {Hensen}}, \bibinfo {author} {\bibfnamefont
  {R.~C.~C.}\ \bibnamefont {Leon}}, \bibinfo {author} {\bibfnamefont {M.~A.}\
  \bibnamefont {Fogarty}}, \bibinfo {author} {\bibfnamefont {J.~C.~C.}\
  \bibnamefont {Hwang}}, \bibinfo {author} {\bibfnamefont {F.~E.}\ \bibnamefont
  {Hudson}}, \bibinfo {author} {\bibfnamefont {K.~M.}\ \bibnamefont {Itoh}},
  \bibinfo {author} {\bibfnamefont {A.}~\bibnamefont {Morello}}, \bibinfo
  {author} {\bibfnamefont {A.}~\bibnamefont {Laucht}}, \ and\ \bibinfo {author}
  {\bibfnamefont {A.~S.}\ \bibnamefont {Dzurak}},\ }\href
  {http://arxiv.org/abs/1805.05027} {\ }\Eprint
  {http://arxiv.org/abs/1805.05027} {arXiv:1805.05027} \BibitemShut {NoStop}%
\bibitem [{\citenamefont {Kelly}\ \emph {et~al.}(2015)\citenamefont {Kelly},
  \citenamefont {Barends}, \citenamefont {Fowler}, \citenamefont {Megrant},
  \citenamefont {Jeffrey}, \citenamefont {White}, \citenamefont {Sank},
  \citenamefont {Mutus}, \citenamefont {Campbell}, \citenamefont {Chen},
  \citenamefont {Chen}, \citenamefont {Chiaro}, \citenamefont {Dunsworth},
  \citenamefont {Hoi}, \citenamefont {Neill}, \citenamefont {Malley},
  \citenamefont {Quintana}, \citenamefont {Roushan}, \citenamefont
  {Vainsencher}, \citenamefont {Wenner}, \citenamefont {Cleland},\ and\
  \citenamefont {Martinis}}]{Kelly2015}%
  \BibitemOpen
  \bibfield  {author} {\bibinfo {author} {\bibfnamefont {J.}~\bibnamefont
  {Kelly}}, \bibinfo {author} {\bibfnamefont {R.}~\bibnamefont {Barends}},
  \bibinfo {author} {\bibfnamefont {A.~G.}\ \bibnamefont {Fowler}}, \bibinfo
  {author} {\bibfnamefont {A.}~\bibnamefont {Megrant}}, \bibinfo {author}
  {\bibfnamefont {E.}~\bibnamefont {Jeffrey}}, \bibinfo {author} {\bibfnamefont
  {T.~C.}\ \bibnamefont {White}}, \bibinfo {author} {\bibfnamefont
  {D.}~\bibnamefont {Sank}}, \bibinfo {author} {\bibfnamefont {J.~Y.}\
  \bibnamefont {Mutus}}, \bibinfo {author} {\bibfnamefont {B.}~\bibnamefont
  {Campbell}}, \bibinfo {author} {\bibfnamefont {Y.}~\bibnamefont {Chen}},
  \bibinfo {author} {\bibfnamefont {Z.}~\bibnamefont {Chen}}, \bibinfo {author}
  {\bibfnamefont {B.}~\bibnamefont {Chiaro}}, \bibinfo {author} {\bibfnamefont
  {A.}~\bibnamefont {Dunsworth}}, \bibinfo {author} {\bibfnamefont
  {I.}~\bibnamefont {Hoi}}, \bibinfo {author} {\bibfnamefont {C.}~\bibnamefont
  {Neill}}, \bibinfo {author} {\bibfnamefont {P.~J. J.~O.}\ \bibnamefont
  {Malley}}, \bibinfo {author} {\bibfnamefont {C.}~\bibnamefont {Quintana}},
  \bibinfo {author} {\bibfnamefont {P.}~\bibnamefont {Roushan}}, \bibinfo
  {author} {\bibfnamefont {A.}~\bibnamefont {Vainsencher}}, \bibinfo {author}
  {\bibfnamefont {J.}~\bibnamefont {Wenner}}, \bibinfo {author} {\bibfnamefont
  {A.~N.}\ \bibnamefont {Cleland}}, \ and\ \bibinfo {author} {\bibfnamefont
  {J.~M.}\ \bibnamefont {Martinis}},\ }\href {\doibase 10.1038/nature14270}
  {\bibfield  {journal} {\bibinfo  {journal} {Nature}\ }\textbf {\bibinfo
  {volume} {519}},\ \bibinfo {pages} {66} (\bibinfo {year} {2015})}\BibitemShut
  {NoStop}%
\bibitem [{\citenamefont {Barends}\ \emph {et~al.}(2014)\citenamefont
  {Barends}, \citenamefont {Kelly}, \citenamefont {Megrant}, \citenamefont
  {Veitia}, \citenamefont {Sank}, \citenamefont {Jeffrey}, \citenamefont
  {White}, \citenamefont {Mutus}, \citenamefont {Fowler}, \citenamefont
  {Campbell}, \citenamefont {Chen}, \citenamefont {Chen}, \citenamefont
  {Chiaro}, \citenamefont {Dunsworth}, \citenamefont {Neill}, \citenamefont
  {O'Malley}, \citenamefont {Roushan}, \citenamefont {Vainsencher},
  \citenamefont {Wenner}, \citenamefont {Korotkov}, \citenamefont {Cleland},\
  and\ \citenamefont {Martinis}}]{Barends2014}%
  \BibitemOpen
  \bibfield  {author} {\bibinfo {author} {\bibfnamefont {R.}~\bibnamefont
  {Barends}}, \bibinfo {author} {\bibfnamefont {J.}~\bibnamefont {Kelly}},
  \bibinfo {author} {\bibfnamefont {A.}~\bibnamefont {Megrant}}, \bibinfo
  {author} {\bibfnamefont {A.}~\bibnamefont {Veitia}}, \bibinfo {author}
  {\bibfnamefont {D.}~\bibnamefont {Sank}}, \bibinfo {author} {\bibfnamefont
  {E.}~\bibnamefont {Jeffrey}}, \bibinfo {author} {\bibfnamefont {T.~C.}\
  \bibnamefont {White}}, \bibinfo {author} {\bibfnamefont {J.}~\bibnamefont
  {Mutus}}, \bibinfo {author} {\bibfnamefont {A.~G.}\ \bibnamefont {Fowler}},
  \bibinfo {author} {\bibfnamefont {B.}~\bibnamefont {Campbell}}, \bibinfo
  {author} {\bibfnamefont {Y.}~\bibnamefont {Chen}}, \bibinfo {author}
  {\bibfnamefont {Z.}~\bibnamefont {Chen}}, \bibinfo {author} {\bibfnamefont
  {B.}~\bibnamefont {Chiaro}}, \bibinfo {author} {\bibfnamefont
  {A.}~\bibnamefont {Dunsworth}}, \bibinfo {author} {\bibfnamefont
  {C.}~\bibnamefont {Neill}}, \bibinfo {author} {\bibfnamefont
  {P.}~\bibnamefont {O'Malley}}, \bibinfo {author} {\bibfnamefont
  {P.}~\bibnamefont {Roushan}}, \bibinfo {author} {\bibfnamefont
  {A.}~\bibnamefont {Vainsencher}}, \bibinfo {author} {\bibfnamefont
  {J.}~\bibnamefont {Wenner}}, \bibinfo {author} {\bibfnamefont {A.~N.}\
  \bibnamefont {Korotkov}}, \bibinfo {author} {\bibfnamefont {A.~N.}\
  \bibnamefont {Cleland}}, \ and\ \bibinfo {author} {\bibfnamefont {J.~M.}\
  \bibnamefont {Martinis}},\ }\href {\doibase 10.1038/nature13171} {\bibfield
  {journal} {\bibinfo  {journal} {Nature}\ }\textbf {\bibinfo {volume} {508}},\
  \bibinfo {pages} {500} (\bibinfo {year} {2014})}\BibitemShut {NoStop}%
\bibitem [{\citenamefont {Fried}\ \emph {et~al.}()\citenamefont {Fried},
  \citenamefont {Hong}, \citenamefont {Karalekas}, \citenamefont {Osborn},
  \citenamefont {Papageorge}, \citenamefont {Peterson}, \citenamefont
  {Prawiroatmodjo}, \citenamefont {Rubin}, \citenamefont {Ryan}, \citenamefont
  {Scarabelli}, \citenamefont {Scheer}, \citenamefont {Sete}, \citenamefont
  {Sivarajah}, \citenamefont {Smith}, \citenamefont {Staley}, \citenamefont
  {Tezak}, \citenamefont {Zeng}, \citenamefont {Hudson}, \citenamefont
  {Johnson}, \citenamefont {Reagor}, \citenamefont {Silva},\ and\ \citenamefont
  {Rigetti}}]{Fried2017}%
  \BibitemOpen
  \bibfield  {author} {\bibinfo {author} {\bibfnamefont {S.}~\bibnamefont
  {Fried}}, \bibinfo {author} {\bibfnamefont {S.}~\bibnamefont {Hong}},
  \bibinfo {author} {\bibfnamefont {P.}~\bibnamefont {Karalekas}}, \bibinfo
  {author} {\bibfnamefont {C.~B.}\ \bibnamefont {Osborn}}, \bibinfo {author}
  {\bibfnamefont {A.}~\bibnamefont {Papageorge}}, \bibinfo {author}
  {\bibfnamefont {E.~C.}\ \bibnamefont {Peterson}}, \bibinfo {author}
  {\bibfnamefont {G.}~\bibnamefont {Prawiroatmodjo}}, \bibinfo {author}
  {\bibfnamefont {N.}~\bibnamefont {Rubin}}, \bibinfo {author} {\bibfnamefont
  {C.~A.}\ \bibnamefont {Ryan}}, \bibinfo {author} {\bibfnamefont
  {D.}~\bibnamefont {Scarabelli}}, \bibinfo {author} {\bibfnamefont
  {M.}~\bibnamefont {Scheer}}, \bibinfo {author} {\bibfnamefont {E.~A.}\
  \bibnamefont {Sete}}, \bibinfo {author} {\bibfnamefont {P.}~\bibnamefont
  {Sivarajah}}, \bibinfo {author} {\bibfnamefont {R.~S.}\ \bibnamefont
  {Smith}}, \bibinfo {author} {\bibfnamefont {A.}~\bibnamefont {Staley}},
  \bibinfo {author} {\bibfnamefont {N.}~\bibnamefont {Tezak}}, \bibinfo
  {author} {\bibfnamefont {W.~J.}\ \bibnamefont {Zeng}}, \bibinfo {author}
  {\bibfnamefont {A.}~\bibnamefont {Hudson}}, \bibinfo {author} {\bibfnamefont
  {B.~R.}\ \bibnamefont {Johnson}}, \bibinfo {author} {\bibfnamefont
  {M.}~\bibnamefont {Reagor}}, \bibinfo {author} {\bibfnamefont {M.~P.}\
  \bibnamefont {Silva}}, \ and\ \bibinfo {author} {\bibfnamefont
  {C.}~\bibnamefont {Rigetti}},\ }\href@noop {} {\ }\Eprint
  {http://arxiv.org/abs/1712.05771v1} {arXiv:1712.05771v1} \BibitemShut
  {NoStop}%
\bibitem [{\citenamefont {Klinovaja}\ \emph {et~al.}(2012)\citenamefont
  {Klinovaja}, \citenamefont {Stepanenko}, \citenamefont {Halperin},\ and\
  \citenamefont {Loss}}]{Klinovaja2012}%
  \BibitemOpen
  \bibfield  {author} {\bibinfo {author} {\bibfnamefont {J.}~\bibnamefont
  {Klinovaja}}, \bibinfo {author} {\bibfnamefont {D.}~\bibnamefont
  {Stepanenko}}, \bibinfo {author} {\bibfnamefont {B.~I.}\ \bibnamefont
  {Halperin}}, \ and\ \bibinfo {author} {\bibfnamefont {D.}~\bibnamefont
  {Loss}},\ }\href {\doibase 10.1103/PhysRevB.86.085423} {\bibfield  {journal}
  {\bibinfo  {journal} {Phys. Rev. B}\ }\textbf {\bibinfo {volume} {86}},\
  \bibinfo {pages} {085423} (\bibinfo {year} {2012})}\BibitemShut {NoStop}%
\bibitem [{\citenamefont {Wardrop}\ and\ \citenamefont
  {Doherty}(2014)}]{Wardrop2014}%
  \BibitemOpen
  \bibfield  {author} {\bibinfo {author} {\bibfnamefont {M.~P.}\ \bibnamefont
  {Wardrop}}\ and\ \bibinfo {author} {\bibfnamefont {A.~C.}\ \bibnamefont
  {Doherty}},\ }\href {\doibase 10.1103/PhysRevB.90.045418} {\bibfield
  {journal} {\bibinfo  {journal} {Phys. Rev. B}\ }\textbf {\bibinfo {volume}
  {90}},\ \bibinfo {pages} {045418} (\bibinfo {year} {2014})}\BibitemShut
  {NoStop}%
\bibitem [{\citenamefont {Li}\ \emph {et~al.}(2012)\citenamefont {Li},
  \citenamefont {Hu},\ and\ \citenamefont {You}}]{Li2012}%
  \BibitemOpen
  \bibfield  {author} {\bibinfo {author} {\bibfnamefont {R.}~\bibnamefont
  {Li}}, \bibinfo {author} {\bibfnamefont {X.}~\bibnamefont {Hu}}, \ and\
  \bibinfo {author} {\bibfnamefont {J.~Q.}\ \bibnamefont {You}},\ }\href
  {\doibase 10.1103/PhysRevB.86.205306} {\bibfield  {journal} {\bibinfo
  {journal} {Phys. Rev. B}\ }\textbf {\bibinfo {volume} {86}},\ \bibinfo
  {pages} {205306} (\bibinfo {year} {2012})}\BibitemShut {NoStop}%
\bibitem [{\citenamefont {Levy}(2002)}]{Levy2002}%
  \BibitemOpen
  \bibfield  {author} {\bibinfo {author} {\bibfnamefont {J.}~\bibnamefont
  {Levy}},\ }\href {\doibase 10.1103/PhysRevLett.89.147902} {\bibfield
  {journal} {\bibinfo  {journal} {Phys. Rev. Lett.}\ }\textbf {\bibinfo
  {volume} {89}},\ \bibinfo {pages} {147902} (\bibinfo {year}
  {2002})}\BibitemShut {NoStop}%
\bibitem [{\citenamefont {Mehl}\ \emph {et~al.}(2014)\citenamefont {Mehl},
  \citenamefont {Bluhm},\ and\ \citenamefont {DiVincenzo}}]{Mehl2014b}%
  \BibitemOpen
  \bibfield  {author} {\bibinfo {author} {\bibfnamefont {S.}~\bibnamefont
  {Mehl}}, \bibinfo {author} {\bibfnamefont {H.}~\bibnamefont {Bluhm}}, \ and\
  \bibinfo {author} {\bibfnamefont {D.~P.}\ \bibnamefont {DiVincenzo}},\ }\href
  {\doibase 10.1103/PhysRevB.90.045404} {\bibfield  {journal} {\bibinfo
  {journal} {Phys. Rev. B}\ }\textbf {\bibinfo {volume} {90}},\ \bibinfo
  {pages} {045404} (\bibinfo {year} {2014})}\BibitemShut {NoStop}%
\bibitem [{\citenamefont {Cerfontaine}\ \emph
  {et~al.}(2019{\natexlab{a}})\citenamefont {Cerfontaine}, \citenamefont
  {Botzem}, \citenamefont {Ritzmann}, \citenamefont {Humpohl}, \citenamefont
  {Ludwig}, \citenamefont {Schuh}, \citenamefont {Bougeard}, \citenamefont
  {Wieck},\ and\ \citenamefont {Bluhm}}]{Cerfontaine2019ex}%
  \BibitemOpen
  \bibfield  {author} {\bibinfo {author} {\bibfnamefont {P.}~\bibnamefont
  {Cerfontaine}}, \bibinfo {author} {\bibfnamefont {T.}~\bibnamefont {Botzem}},
  \bibinfo {author} {\bibfnamefont {J.}~\bibnamefont {Ritzmann}}, \bibinfo
  {author} {\bibfnamefont {S.~S.}\ \bibnamefont {Humpohl}}, \bibinfo {author}
  {\bibfnamefont {A.}~\bibnamefont {Ludwig}}, \bibinfo {author} {\bibfnamefont
  {D.}~\bibnamefont {Schuh}}, \bibinfo {author} {\bibfnamefont
  {D.}~\bibnamefont {Bougeard}}, \bibinfo {author} {\bibfnamefont {A.~D.}\
  \bibnamefont {Wieck}}, \ and\ \bibinfo {author} {\bibfnamefont
  {H.}~\bibnamefont {Bluhm}},\ }\href@noop {} {\bibfield  {journal} {\bibinfo
  {journal} {arXiv:1906.06169}\ } (\bibinfo {year}
  {2019}{\natexlab{a}})}\BibitemShut {NoStop}%
\bibitem [{\citenamefont {Cerfontaine}\ \emph {et~al.}(2014)\citenamefont
  {Cerfontaine}, \citenamefont {Botzem}, \citenamefont {DiVincenzo},\ and\
  \citenamefont {Bluhm}}]{Cerfontaine2014}%
  \BibitemOpen
  \bibfield  {author} {\bibinfo {author} {\bibfnamefont {P.}~\bibnamefont
  {Cerfontaine}}, \bibinfo {author} {\bibfnamefont {T.}~\bibnamefont {Botzem}},
  \bibinfo {author} {\bibfnamefont {D.~P.}\ \bibnamefont {DiVincenzo}}, \ and\
  \bibinfo {author} {\bibfnamefont {H.}~\bibnamefont {Bluhm}},\ }\href
  {\doibase 10.1103/PhysRevLett.113.150501} {\bibfield  {journal} {\bibinfo
  {journal} {Phys. Rev. Lett.}\ }\textbf {\bibinfo {volume} {113}},\ \bibinfo
  {pages} {150501} (\bibinfo {year} {2014})}\BibitemShut {NoStop}%
\bibitem [{\citenamefont {Shulman}\ \emph {et~al.}(2012)\citenamefont
  {Shulman}, \citenamefont {Dial}, \citenamefont {Harvey}, \citenamefont
  {Bluhm}, \citenamefont {Umansky},\ and\ \citenamefont
  {Yacoby}}]{Shulman2012}%
  \BibitemOpen
  \bibfield  {author} {\bibinfo {author} {\bibfnamefont {M.~D.}\ \bibnamefont
  {Shulman}}, \bibinfo {author} {\bibfnamefont {O.~E.}\ \bibnamefont {Dial}},
  \bibinfo {author} {\bibfnamefont {S.~P.}\ \bibnamefont {Harvey}}, \bibinfo
  {author} {\bibfnamefont {H.}~\bibnamefont {Bluhm}}, \bibinfo {author}
  {\bibfnamefont {V.}~\bibnamefont {Umansky}}, \ and\ \bibinfo {author}
  {\bibfnamefont {A.}~\bibnamefont {Yacoby}},\ }\href {\doibase
  10.1126/science.1217692} {\bibfield  {journal} {\bibinfo  {journal}
  {Science}\ }\textbf {\bibinfo {volume} {336}},\ \bibinfo {pages} {202}
  (\bibinfo {year} {2012})}\BibitemShut {NoStop}%
\bibitem [{\citenamefont {Nichol}\ \emph {et~al.}(2017)\citenamefont {Nichol},
  \citenamefont {Orona}, \citenamefont {Harvey}, \citenamefont {Fallahi},
  \citenamefont {Gardner},\ and\ \citenamefont {Manfra}}]{Nichol2017}%
  \BibitemOpen
  \bibfield  {author} {\bibinfo {author} {\bibfnamefont {J.~M.}\ \bibnamefont
  {Nichol}}, \bibinfo {author} {\bibfnamefont {L.~A.}\ \bibnamefont {Orona}},
  \bibinfo {author} {\bibfnamefont {S.~P.}\ \bibnamefont {Harvey}}, \bibinfo
  {author} {\bibfnamefont {S.}~\bibnamefont {Fallahi}}, \bibinfo {author}
  {\bibfnamefont {G.~C.}\ \bibnamefont {Gardner}}, \ and\ \bibinfo {author}
  {\bibfnamefont {A.}~\bibnamefont {Manfra}, \bibfnamefont {Michael
  J.and~Yacoby}},\ }\href {http://dx.doi.org/10.1038/s41534-016-0003-1}
  {\bibfield  {journal} {\bibinfo  {journal} {npj Quantum Inf.}\ }\textbf
  {\bibinfo {volume} {3}} (\bibinfo {year} {2017})}\BibitemShut {NoStop}%
\bibitem [{\citenamefont {Buterakos}\ \emph
  {et~al.}(2018{\natexlab{a}})\citenamefont {Buterakos}, \citenamefont
  {Throckmorton},\ and\ \citenamefont {{Das Sarma}}}]{Buterakos2018}%
  \BibitemOpen
  \bibfield  {author} {\bibinfo {author} {\bibfnamefont {D.}~\bibnamefont
  {Buterakos}}, \bibinfo {author} {\bibfnamefont {R.~E.}\ \bibnamefont
  {Throckmorton}}, \ and\ \bibinfo {author} {\bibfnamefont {S.}~\bibnamefont
  {{Das Sarma}}},\ }\href {\doibase 10.1103/PhysRevB.98.035406} {\bibfield
  {journal} {\bibinfo  {journal} {Phys. Rev. B}\ }\textbf {\bibinfo {volume}
  {98}},\ \bibinfo {pages} {035406} (\bibinfo {year}
  {2018}{\natexlab{a}})}\BibitemShut {NoStop}%
\bibitem [{\citenamefont {Buterakos}\ \emph
  {et~al.}(2018{\natexlab{b}})\citenamefont {Buterakos}, \citenamefont
  {Throckmorton},\ and\ \citenamefont {{Das Sarma}}}]{Buterakos2018a}%
  \BibitemOpen
  \bibfield  {author} {\bibinfo {author} {\bibfnamefont {D.}~\bibnamefont
  {Buterakos}}, \bibinfo {author} {\bibfnamefont {R.~E.}\ \bibnamefont
  {Throckmorton}}, \ and\ \bibinfo {author} {\bibfnamefont {S.}~\bibnamefont
  {{Das Sarma}}},\ }\href {\doibase 10.1103/PhysRevB.97.045431} {\bibfield
  {journal} {\bibinfo  {journal} {Physical Review B}\ }\textbf {\bibinfo
  {volume} {97}},\ \bibinfo {pages} {045431} (\bibinfo {year}
  {2018}{\natexlab{b}})}\BibitemShut {NoStop}%
\bibitem [{\citenamefont {Petta}\ \emph {et~al.}(2005)\citenamefont {Petta},
  \citenamefont {Johnson}, \citenamefont {Taylor}, \citenamefont {Laird},
  \citenamefont {Yacoby}, \citenamefont {Lukin}, \citenamefont {Marcus},
  \citenamefont {Hanson},\ and\ \citenamefont {Gossard}}]{Petta2005}%
  \BibitemOpen
  \bibfield  {author} {\bibinfo {author} {\bibfnamefont {J.~R.}\ \bibnamefont
  {Petta}}, \bibinfo {author} {\bibfnamefont {A.~C.}\ \bibnamefont {Johnson}},
  \bibinfo {author} {\bibfnamefont {J.~M.}\ \bibnamefont {Taylor}}, \bibinfo
  {author} {\bibfnamefont {E.~A.}\ \bibnamefont {Laird}}, \bibinfo {author}
  {\bibfnamefont {A.}~\bibnamefont {Yacoby}}, \bibinfo {author} {\bibfnamefont
  {M.~D.}\ \bibnamefont {Lukin}}, \bibinfo {author} {\bibfnamefont {C.~M.}\
  \bibnamefont {Marcus}}, \bibinfo {author} {\bibfnamefont {M.~P.}\
  \bibnamefont {Hanson}}, \ and\ \bibinfo {author} {\bibfnamefont {A.~C.}\
  \bibnamefont {Gossard}},\ }\href {\doibase 10.1126/science.1116955}
  {\bibfield  {journal} {\bibinfo  {journal} {Science}\ }\textbf {\bibinfo
  {volume} {309}},\ \bibinfo {pages} {2180} (\bibinfo {year}
  {2005})}\BibitemShut {NoStop}%
\bibitem [{\citenamefont {Dial}\ \emph {et~al.}(2013)\citenamefont {Dial},
  \citenamefont {Shulman}, \citenamefont {Harvey}, \citenamefont {Bluhm},
  \citenamefont {Umansky},\ and\ \citenamefont {Yacoby}}]{Dial2013}%
  \BibitemOpen
  \bibfield  {author} {\bibinfo {author} {\bibfnamefont {O.~E.}\ \bibnamefont
  {Dial}}, \bibinfo {author} {\bibfnamefont {M.~D.}\ \bibnamefont {Shulman}},
  \bibinfo {author} {\bibfnamefont {S.~P.}\ \bibnamefont {Harvey}}, \bibinfo
  {author} {\bibfnamefont {H.}~\bibnamefont {Bluhm}}, \bibinfo {author}
  {\bibfnamefont {V.}~\bibnamefont {Umansky}}, \ and\ \bibinfo {author}
  {\bibfnamefont {A.}~\bibnamefont {Yacoby}},\ }\href {\doibase
  10.1103/PhysRevLett.110.146804} {\bibfield  {journal} {\bibinfo  {journal}
  {Phys. Rev. Lett.}\ }\textbf {\bibinfo {volume} {110}},\ \bibinfo {pages}
  {146804} (\bibinfo {year} {2013})}\BibitemShut {NoStop}%
\bibitem [{\citenamefont {Reed}\ \emph {et~al.}(2016)\citenamefont {Reed},
  \citenamefont {Maune}, \citenamefont {Andrews}, \citenamefont {Borselli},
  \citenamefont {Eng}, \citenamefont {Jura}, \citenamefont {Kiselev},
  \citenamefont {Ladd}, \citenamefont {Merkel}, \citenamefont {Milosavljevic},
  \citenamefont {Pritchett}, \citenamefont {Rakher}, \citenamefont {Ross},
  \citenamefont {Schmitz}, \citenamefont {Smith}, \citenamefont {Wright},
  \citenamefont {Gyure},\ and\ \citenamefont {Hunter}}]{Reed2016}%
  \BibitemOpen
  \bibfield  {author} {\bibinfo {author} {\bibfnamefont {M.~D.}\ \bibnamefont
  {Reed}}, \bibinfo {author} {\bibfnamefont {B.~M.}\ \bibnamefont {Maune}},
  \bibinfo {author} {\bibfnamefont {R.~W.}\ \bibnamefont {Andrews}}, \bibinfo
  {author} {\bibfnamefont {M.~G.}\ \bibnamefont {Borselli}}, \bibinfo {author}
  {\bibfnamefont {K.}~\bibnamefont {Eng}}, \bibinfo {author} {\bibfnamefont
  {M.~P.}\ \bibnamefont {Jura}}, \bibinfo {author} {\bibfnamefont {A.~A.}\
  \bibnamefont {Kiselev}}, \bibinfo {author} {\bibfnamefont {T.~D.}\
  \bibnamefont {Ladd}}, \bibinfo {author} {\bibfnamefont {S.~T.}\ \bibnamefont
  {Merkel}}, \bibinfo {author} {\bibfnamefont {I.}~\bibnamefont
  {Milosavljevic}}, \bibinfo {author} {\bibfnamefont {E.~J.}\ \bibnamefont
  {Pritchett}}, \bibinfo {author} {\bibfnamefont {M.~T.}\ \bibnamefont
  {Rakher}}, \bibinfo {author} {\bibfnamefont {R.~S.}\ \bibnamefont {Ross}},
  \bibinfo {author} {\bibfnamefont {A.~E.}\ \bibnamefont {Schmitz}}, \bibinfo
  {author} {\bibfnamefont {A.}~\bibnamefont {Smith}}, \bibinfo {author}
  {\bibfnamefont {J.~A.}\ \bibnamefont {Wright}}, \bibinfo {author}
  {\bibfnamefont {M.~F.}\ \bibnamefont {Gyure}}, \ and\ \bibinfo {author}
  {\bibfnamefont {A.~T.}\ \bibnamefont {Hunter}},\ }\href {\doibase
  10.1103/PhysRevLett.116.110402} {\bibfield  {journal} {\bibinfo  {journal}
  {Phys. Rev. Lett.}\ }\textbf {\bibinfo {volume} {116}},\ \bibinfo {pages}
  {110402} (\bibinfo {year} {2016})}\BibitemShut {NoStop}%
\bibitem [{\citenamefont {Martins}\ \emph {et~al.}(2016)\citenamefont
  {Martins}, \citenamefont {Malinowski}, \citenamefont {Nissen}, \citenamefont
  {Barnes}, \citenamefont {Fallahi}, \citenamefont {Gardner}, \citenamefont
  {Manfra}, \citenamefont {Marcus},\ and\ \citenamefont
  {Kuemmeth}}]{Martins2016}%
  \BibitemOpen
  \bibfield  {author} {\bibinfo {author} {\bibfnamefont {F.}~\bibnamefont
  {Martins}}, \bibinfo {author} {\bibfnamefont {F.~K.}\ \bibnamefont
  {Malinowski}}, \bibinfo {author} {\bibfnamefont {P.~D.}\ \bibnamefont
  {Nissen}}, \bibinfo {author} {\bibfnamefont {E.}~\bibnamefont {Barnes}},
  \bibinfo {author} {\bibfnamefont {S.}~\bibnamefont {Fallahi}}, \bibinfo
  {author} {\bibfnamefont {G.~C.}\ \bibnamefont {Gardner}}, \bibinfo {author}
  {\bibfnamefont {M.~J.}\ \bibnamefont {Manfra}}, \bibinfo {author}
  {\bibfnamefont {C.~M.}\ \bibnamefont {Marcus}}, \ and\ \bibinfo {author}
  {\bibfnamefont {F.}~\bibnamefont {Kuemmeth}},\ }\href {\doibase
  10.1103/PhysRevLett.116.116801} {\bibfield  {journal} {\bibinfo  {journal}
  {Phys. Rev. Lett.}\ }\textbf {\bibinfo {volume} {116}},\ \bibinfo {pages}
  {116801} (\bibinfo {year} {2016})}\BibitemShut {NoStop}%
\bibitem [{\citenamefont {Wu}\ \emph {et~al.}(2014)\citenamefont {Wu},
  \citenamefont {Ward}, \citenamefont {Prance}, \citenamefont {Kim},
  \citenamefont {Gamble}, \citenamefont {Mohr}, \citenamefont {Shi},
  \citenamefont {Savage}, \citenamefont {Lagally}, \citenamefont {Friesen},
  \citenamefont {Coppersmith},\ and\ \citenamefont {Eriksson}}]{Wu2014}%
  \BibitemOpen
  \bibfield  {author} {\bibinfo {author} {\bibfnamefont {X.}~\bibnamefont
  {Wu}}, \bibinfo {author} {\bibfnamefont {D.~R.}\ \bibnamefont {Ward}},
  \bibinfo {author} {\bibfnamefont {J.~R.}\ \bibnamefont {Prance}}, \bibinfo
  {author} {\bibfnamefont {D.}~\bibnamefont {Kim}}, \bibinfo {author}
  {\bibfnamefont {J.~K.}\ \bibnamefont {Gamble}}, \bibinfo {author}
  {\bibfnamefont {R.~T.}\ \bibnamefont {Mohr}}, \bibinfo {author}
  {\bibfnamefont {Z.}~\bibnamefont {Shi}}, \bibinfo {author} {\bibfnamefont
  {D.~E.}\ \bibnamefont {Savage}}, \bibinfo {author} {\bibfnamefont {M.~G.}\
  \bibnamefont {Lagally}}, \bibinfo {author} {\bibfnamefont {M.}~\bibnamefont
  {Friesen}}, \bibinfo {author} {\bibfnamefont {S.~N.}\ \bibnamefont
  {Coppersmith}}, \ and\ \bibinfo {author} {\bibfnamefont {M.~A.}\ \bibnamefont
  {Eriksson}},\ }\href {\doibase 10.1073/pnas.1412230111} {\bibfield  {journal}
  {\bibinfo  {journal} {PNAS}\ }\textbf {\bibinfo {volume} {111}},\ \bibinfo
  {pages} {11938} (\bibinfo {year} {2014})}\BibitemShut {NoStop}%
\bibitem [{\citenamefont {Veldhorst}\ \emph {et~al.}(2014)\citenamefont
  {Veldhorst}, \citenamefont {Hwang}, \citenamefont {Yang}, \citenamefont
  {Leenstra}, \citenamefont {de~Ronde}, \citenamefont {Dehollain},
  \citenamefont {Muhonen}, \citenamefont {Hudson}, \citenamefont {Itoh},
  \citenamefont {Morello},\ and\ \citenamefont {Dzurak}}]{Veldhorst2014}%
  \BibitemOpen
  \bibfield  {author} {\bibinfo {author} {\bibfnamefont {M.}~\bibnamefont
  {Veldhorst}}, \bibinfo {author} {\bibfnamefont {J.~C.~C.}\ \bibnamefont
  {Hwang}}, \bibinfo {author} {\bibfnamefont {C.~H.}\ \bibnamefont {Yang}},
  \bibinfo {author} {\bibfnamefont {A.~W.}\ \bibnamefont {Leenstra}}, \bibinfo
  {author} {\bibfnamefont {B.}~\bibnamefont {de~Ronde}}, \bibinfo {author}
  {\bibfnamefont {J.~P.}\ \bibnamefont {Dehollain}}, \bibinfo {author}
  {\bibfnamefont {J.~T.}\ \bibnamefont {Muhonen}}, \bibinfo {author}
  {\bibfnamefont {F.~E.}\ \bibnamefont {Hudson}}, \bibinfo {author}
  {\bibfnamefont {K.~M.}\ \bibnamefont {Itoh}}, \bibinfo {author}
  {\bibfnamefont {A.}~\bibnamefont {Morello}}, \ and\ \bibinfo {author}
  {\bibfnamefont {A.~S.}\ \bibnamefont {Dzurak}},\ }\href {\doibase
  10.1038/nnano.2014.216} {\bibfield  {journal} {\bibinfo  {journal} {Nat.
  Nanotechnol.}\ }\textbf {\bibinfo {volume} {9}},\ \bibinfo {pages} {981}
  (\bibinfo {year} {2014})}\BibitemShut {NoStop}%
\bibitem [{\citenamefont {Bluhm}\ \emph {et~al.}(2010)\citenamefont {Bluhm},
  \citenamefont {Foletti}, \citenamefont {Mahalu}, \citenamefont {Umansky},\
  and\ \citenamefont {Yacoby}}]{Bluhm2010}%
  \BibitemOpen
  \bibfield  {author} {\bibinfo {author} {\bibfnamefont {H.}~\bibnamefont
  {Bluhm}}, \bibinfo {author} {\bibfnamefont {S.}~\bibnamefont {Foletti}},
  \bibinfo {author} {\bibfnamefont {D.}~\bibnamefont {Mahalu}}, \bibinfo
  {author} {\bibfnamefont {V.}~\bibnamefont {Umansky}}, \ and\ \bibinfo
  {author} {\bibfnamefont {A.}~\bibnamefont {Yacoby}},\ }\href {\doibase
  10.1103/PhysRevLett.105.216803} {\bibfield  {journal} {\bibinfo  {journal}
  {Phys. Rev. Lett.}\ }\textbf {\bibinfo {volume} {105}},\ \bibinfo {pages}
  {216803} (\bibinfo {year} {2010})}\BibitemShut {NoStop}%
\bibitem [{\citenamefont {Reilly}\ \emph {et~al.}(2008)\citenamefont {Reilly},
  \citenamefont {Taylor}, \citenamefont {Laird}, \citenamefont {Petta},
  \citenamefont {Marcus}, \citenamefont {Hanson},\ and\ \citenamefont
  {Gossard}}]{Reilly2008}%
  \BibitemOpen
  \bibfield  {author} {\bibinfo {author} {\bibfnamefont {D.~J.}\ \bibnamefont
  {Reilly}}, \bibinfo {author} {\bibfnamefont {J.~M.}\ \bibnamefont {Taylor}},
  \bibinfo {author} {\bibfnamefont {E.~A.}\ \bibnamefont {Laird}}, \bibinfo
  {author} {\bibfnamefont {J.~R.}\ \bibnamefont {Petta}}, \bibinfo {author}
  {\bibfnamefont {C.~M.}\ \bibnamefont {Marcus}}, \bibinfo {author}
  {\bibfnamefont {M.~P.}\ \bibnamefont {Hanson}}, \ and\ \bibinfo {author}
  {\bibfnamefont {A.~C.}\ \bibnamefont {Gossard}},\ }\href {\doibase
  10.1103/PhysRevLett.101.236803} {\bibfield  {journal} {\bibinfo  {journal}
  {Phys. Rev. Lett.}\ }\textbf {\bibinfo {volume} {101}},\ \bibinfo {pages}
  {236803} (\bibinfo {year} {2008})}\BibitemShut {NoStop}%
\bibitem [{\citenamefont {Barthel}\ \emph {et~al.}(2009)\citenamefont
  {Barthel}, \citenamefont {Reilly}, \citenamefont {Marcus}, \citenamefont
  {Hanson},\ and\ \citenamefont {Gossard}}]{Barthel2009}%
  \BibitemOpen
  \bibfield  {author} {\bibinfo {author} {\bibfnamefont {C.}~\bibnamefont
  {Barthel}}, \bibinfo {author} {\bibfnamefont {D.J.}~\bibnamefont {Reilly}},
  \bibinfo {author} {\bibfnamefont {C.M.}~\bibnamefont {Marcus}}, \bibinfo
  {author} {\bibfnamefont {M.P.}~\bibnamefont {Hanson}}, \ and\ \bibinfo {author}
  {\bibfnamefont {A.C.}~\bibnamefont {Gossard}},\ }\href {\doibase
  10.1103/PhysRevLett.103.160503} {\bibfield  {journal} {\bibinfo  {journal}
  {Physical Review Letters}\ }\textbf {\bibinfo {volume} {103}},\ \bibinfo
  {pages} {160503} (\bibinfo {year} {2009})}\BibitemShut {NoStop}%
\bibitem [{\citenamefont {Keller}(1975)}]{Keller1975}%
  \BibitemOpen
  \bibfield  {author} {\bibinfo {author} {\bibfnamefont {J.~B.}\ \bibnamefont
  {Keller}},\ }\href@noop {} {\bibfield  {journal} {\bibinfo  {journal} {Math.
  Mag.}\ }\textbf {\bibinfo {volume} {48}},\ \bibinfo {pages} {192} (\bibinfo
  {year} {1975})}\BibitemShut {NoStop}%
\bibitem [{\citenamefont {Nielsen}(2002)}]{Nielsen2002}%
  \BibitemOpen
  \bibfield  {author} {\bibinfo {author} {\bibfnamefont {M.~A.}\ \bibnamefont
  {Nielsen}},\ }\href {\doibase 10.1016/S0375-9601(02)01272-0} {\bibfield
  {journal} {\bibinfo  {journal} {Phys. Lett. A}\ }\textbf {\bibinfo {volume}
  {303}},\ \bibinfo {pages} {249} (\bibinfo {year} {2002})}\BibitemShut
  {NoStop}%
\bibitem [{\citenamefont {Khaneja}\ \emph {et~al.}(2005)\citenamefont
  {Khaneja}, \citenamefont {Reiss}, \citenamefont {Kehlet}, \citenamefont
  {Schulte-Herbr{\"{u}}ggen},\ and\ \citenamefont {Glaser}}]{Khaneja2005}%
  \BibitemOpen
  \bibfield  {author} {\bibinfo {author} {\bibfnamefont {N.}~\bibnamefont
  {Khaneja}}, \bibinfo {author} {\bibfnamefont {T.}~\bibnamefont {Reiss}},
  \bibinfo {author} {\bibfnamefont {C.}~\bibnamefont {Kehlet}}, \bibinfo
  {author} {\bibfnamefont {T.}~\bibnamefont {Schulte-Herbr{\"{u}}ggen}}, \ and\
  \bibinfo {author} {\bibfnamefont {S.~J.}\ \bibnamefont {Glaser}},\ }\href
  {\doibase 10.1016/j.jmr.2004.11.004} {\bibfield  {journal} {\bibinfo
  {journal} {J. Magn. Reson.}\ }\textbf {\bibinfo {volume} {172}},\ \bibinfo
  {pages} {296} (\bibinfo {year} {2005})}\BibitemShut {NoStop}%
\bibitem [{\citenamefont {Kuprov}\ and\ \citenamefont
  {Rodgers}(2009)}]{Kuprov2009}%
  \BibitemOpen
  \bibfield  {author} {\bibinfo {author} {\bibfnamefont {I.}~\bibnamefont
  {Kuprov}}\ and\ \bibinfo {author} {\bibfnamefont {C.~T.}\ \bibnamefont
  {Rodgers}},\ }\href {\doibase 10.1063/1.3267086} {\bibfield  {journal}
  {\bibinfo  {journal} {The Journal of Chemical Physics}\ }\textbf {\bibinfo
  {volume} {131}},\ \bibinfo {pages} {234108} (\bibinfo {year}
  {2009})}\BibitemShut {NoStop}%
\bibitem [{\citenamefont {Floether}\ \emph {et~al.}(2012)\citenamefont
  {Floether}, \citenamefont {de~Fouquieres},\ and\ \citenamefont
  {Schirmer}}]{Floether2011}%
  \BibitemOpen
  \bibfield  {author} {\bibinfo {author} {\bibfnamefont {F.~F.}\ \bibnamefont
  {Floether}}, \bibinfo {author} {\bibfnamefont {P.}~\bibnamefont
  {de~Fouquieres}}, \ and\ \bibinfo {author} {\bibfnamefont {S.~G.}\
  \bibnamefont {Schirmer}},\ }\href {\doibase 10.1088/1367-2630/14/7/073023}
  {\bibfield  {journal} {\bibinfo  {journal} {New Journal of Physics}\ }\textbf
  {\bibinfo {volume} {14}},\ \bibinfo {pages} {073023} (\bibinfo {year}
  {2012})}\BibitemShut {NoStop}%
\bibitem [{\citenamefont {Chan}\ \emph {et~al.}(2018)\citenamefont {Chan},
  \citenamefont {Huang}, \citenamefont {Yang}, \citenamefont {Hwang},
  \citenamefont {Hensen}, \citenamefont {Tanttu}, \citenamefont {Hudson},
  \citenamefont {Itoh}, \citenamefont {Laucht}, \citenamefont {Morello},\ and\
  \citenamefont {Dzurak}}]{Chan2018}%
  \BibitemOpen
  \bibfield  {author} {\bibinfo {author} {\bibfnamefont {K.~W.}\ \bibnamefont
  {Chan}}, \bibinfo {author} {\bibfnamefont {W.}~\bibnamefont {Huang}},
  \bibinfo {author} {\bibfnamefont {C.~H.}\ \bibnamefont {Yang}}, \bibinfo
  {author} {\bibfnamefont {J.~C.~C.}\ \bibnamefont {Hwang}}, \bibinfo {author}
  {\bibfnamefont {B.}~\bibnamefont {Hensen}}, \bibinfo {author} {\bibfnamefont
  {T.}~\bibnamefont {Tanttu}}, \bibinfo {author} {\bibfnamefont {F.~E.}\
  \bibnamefont {Hudson}}, \bibinfo {author} {\bibfnamefont {K.~M.}\
  \bibnamefont {Itoh}}, \bibinfo {author} {\bibfnamefont {A.}~\bibnamefont
  {Laucht}}, \bibinfo {author} {\bibfnamefont {A.}~\bibnamefont {Morello}}, \
  and\ \bibinfo {author} {\bibfnamefont {A.~S.}\ \bibnamefont {Dzurak}},\
  }\href {\doibase 10.1103/PhysRevApplied.10.044017} {\bibfield  {journal}
  {\bibinfo  {journal} {Phys. Rev. Appl.}\ }\textbf {\bibinfo {volume} {10}},\
  \bibinfo {pages} {044017} (\bibinfo {year} {2018})}\BibitemShut {NoStop}%
\bibitem [{\citenamefont {Mi}\ \emph {et~al.}(2018)\citenamefont {Mi},
  \citenamefont {Kohler},\ and\ \citenamefont {Petta}}]{Mi2018}%
  \BibitemOpen
  \bibfield  {author} {\bibinfo {author} {\bibfnamefont {X.}~\bibnamefont
  {Mi}}, \bibinfo {author} {\bibfnamefont {S.}~\bibnamefont {Kohler}}, \ and\
  \bibinfo {author} {\bibfnamefont {J.~R.}\ \bibnamefont {Petta}},\ }\href
  {\doibase 10.1103/PhysRevB.98.161404} {\bibfield  {journal} {\bibinfo
  {journal} {Phys. Rev. B}\ }\textbf {\bibinfo {volume} {98}},\ \bibinfo
  {pages} {161404(R)} (\bibinfo {year} {2018})}\BibitemShut {NoStop}%
\bibitem [{\citenamefont {Eng}\ \emph {et~al.}(2015)\citenamefont {Eng},
  \citenamefont {Ladd}, \citenamefont {Smith}, \citenamefont {Borselli},
  \citenamefont {Kiselev}, \citenamefont {Fong}, \citenamefont {Holabird},
  \citenamefont {Hazard}, \citenamefont {Huang}, \citenamefont {Deelman},
  \citenamefont {Milosavljevic}, \citenamefont {Schmitz}, \citenamefont {Ross},
  \citenamefont {Gyure},\ and\ \citenamefont {Hunter}}]{Eng2015}%
  \BibitemOpen
  \bibfield  {author} {\bibinfo {author} {\bibfnamefont {K.}~\bibnamefont
  {Eng}}, \bibinfo {author} {\bibfnamefont {T.~D.}\ \bibnamefont {Ladd}},
  \bibinfo {author} {\bibfnamefont {A.}~\bibnamefont {Smith}}, \bibinfo
  {author} {\bibfnamefont {M.~G.}\ \bibnamefont {Borselli}}, \bibinfo {author}
  {\bibfnamefont {A.~A.}\ \bibnamefont {Kiselev}}, \bibinfo {author}
  {\bibfnamefont {B.~H.}\ \bibnamefont {Fong}}, \bibinfo {author}
  {\bibfnamefont {K.~S.}\ \bibnamefont {Holabird}}, \bibinfo {author}
  {\bibfnamefont {T.~M.}\ \bibnamefont {Hazard}}, \bibinfo {author}
  {\bibfnamefont {B.}~\bibnamefont {Huang}}, \bibinfo {author} {\bibfnamefont
  {P.~W.}\ \bibnamefont {Deelman}}, \bibinfo {author} {\bibfnamefont
  {I.}~\bibnamefont {Milosavljevic}}, \bibinfo {author} {\bibfnamefont {A.~E.}\
  \bibnamefont {Schmitz}}, \bibinfo {author} {\bibfnamefont {R.~S.}\
  \bibnamefont {Ross}}, \bibinfo {author} {\bibfnamefont {M.~F.}\ \bibnamefont
  {Gyure}}, \ and\ \bibinfo {author} {\bibfnamefont {A.~T.}\ \bibnamefont
  {Hunter}},\ }\href {\doibase 10.1126/sciadv.1500214} {\bibfield  {journal}
  {\bibinfo  {journal} {Sci. Adv.}\ }\textbf {\bibinfo {volume} {1}},\ \bibinfo
  {pages} {e1500214} (\bibinfo {year} {2015})}\BibitemShut {NoStop}%
\bibitem [{\citenamefont {Botzem}\ \emph {et~al.}(2018)\citenamefont {Botzem},
  \citenamefont {Shulman}, \citenamefont {Foletti}, \citenamefont {Harvey},
  \citenamefont {Dial}, \citenamefont {Bethke}, \citenamefont {Cerfontaine},
  \citenamefont {McNeil}, \citenamefont {Mahalu}, \citenamefont {Umansky},
  \citenamefont {Ludwig}, \citenamefont {Wieck}, \citenamefont {Schuh},
  \citenamefont {Bougeard}, \citenamefont {Yacoby},\ and\ \citenamefont
  {Bluhm}}]{Botzem2018}%
  \BibitemOpen
  \bibfield  {author} {\bibinfo {author} {\bibfnamefont {T.}~\bibnamefont
  {Botzem}}, \bibinfo {author} {\bibfnamefont {M.~D.}\ \bibnamefont {Shulman}},
  \bibinfo {author} {\bibfnamefont {S.}~\bibnamefont {Foletti}}, \bibinfo
  {author} {\bibfnamefont {S.~P.}\ \bibnamefont {Harvey}}, \bibinfo {author}
  {\bibfnamefont {O.~E.}\ \bibnamefont {Dial}}, \bibinfo {author}
  {\bibfnamefont {P.}~\bibnamefont {Bethke}}, \bibinfo {author} {\bibfnamefont
  {P.}~\bibnamefont {Cerfontaine}}, \bibinfo {author} {\bibfnamefont
  {R.~P.~G.}\ \bibnamefont {McNeil}}, \bibinfo {author} {\bibfnamefont
  {D.}~\bibnamefont {Mahalu}}, \bibinfo {author} {\bibfnamefont
  {V.}~\bibnamefont {Umansky}}, \bibinfo {author} {\bibfnamefont
  {A.}~\bibnamefont {Ludwig}}, \bibinfo {author} {\bibfnamefont
  {A.}~\bibnamefont {Wieck}}, \bibinfo {author} {\bibfnamefont
  {D.}~\bibnamefont {Schuh}}, \bibinfo {author} {\bibfnamefont
  {D.}~\bibnamefont {Bougeard}}, \bibinfo {author} {\bibfnamefont
  {A.}~\bibnamefont {Yacoby}}, \ and\ \bibinfo {author} {\bibfnamefont
  {H.}~\bibnamefont {Bluhm}},\ }\href {\doibase
  10.1103/PhysRevApplied.10.054026} {\bibfield  {journal} {\bibinfo  {journal}
  {Phys. Rev. Appl.}\ }\textbf {\bibinfo {volume} {10}},\ \bibinfo {pages}
  {054026} (\bibinfo {year} {2018})}\BibitemShut {NoStop}%
\bibitem [{\citenamefont {{Cerfontaine}}\ \emph {et~al.}()\citenamefont
  {{Cerfontaine}}, \citenamefont {{Botzem}}, \citenamefont {{Humpohl}},
  \citenamefont {{Schuh}}, \citenamefont {{Bougeard}},\ and\ \citenamefont
  {{Bluhm}}}]{Cerfontaine2016}%
  \BibitemOpen
  \bibfield  {author} {\bibinfo {author} {\bibfnamefont {P.}~\bibnamefont
  {{Cerfontaine}}}, \bibinfo {author} {\bibfnamefont {T.}~\bibnamefont
  {{Botzem}}}, \bibinfo {author} {\bibfnamefont {S.~S.}\ \bibnamefont
  {{Humpohl}}}, \bibinfo {author} {\bibfnamefont {D.}~\bibnamefont {{Schuh}}},
  \bibinfo {author} {\bibfnamefont {D.}~\bibnamefont {{Bougeard}}}, \ and\
  \bibinfo {author} {\bibfnamefont {H.}~\bibnamefont {{Bluhm}}},\ }\href@noop
  {} {\ }\Eprint {http://arxiv.org/abs/1606.01897} {arXiv:1606.01897}
  \BibitemShut {NoStop}%
\bibitem [{\citenamefont {Cerfontaine}\ \emph
  {et~al.}(2019{\natexlab{b}})\citenamefont {Cerfontaine}, \citenamefont
  {Otten},\ and\ \citenamefont {Bluhm}}]{Cerfontaine2019a}%
  \BibitemOpen
  \bibfield  {author} {\bibinfo {author} {\bibfnamefont {P.}~\bibnamefont
  {Cerfontaine}}, \bibinfo {author} {\bibfnamefont {R.}~\bibnamefont {Otten}},
  \ and\ \bibinfo {author} {\bibfnamefont {H.}~\bibnamefont {Bluhm}},\
  }\href@noop {} {\bibfield  {journal} {\bibinfo  {journal} {arXiv:1906.00950}\
  } (\bibinfo {year} {2019}{\natexlab{b}})}\BibitemShut {NoStop}%
\bibitem [{\citenamefont {Joecker}\ \emph {et~al.}(2019)\citenamefont
  {Joecker}, \citenamefont {Cerfontaine}, \citenamefont {Haupt}, \citenamefont
  {Schreiber}, \citenamefont {Kardyna{\l}},\ and\ \citenamefont
  {Bluhm}}]{Joecker2018}%
  \BibitemOpen
  \bibfield  {author} {\bibinfo {author} {\bibfnamefont {B.}~\bibnamefont
  {Joecker}}, \bibinfo {author} {\bibfnamefont {P.}~\bibnamefont
  {Cerfontaine}}, \bibinfo {author} {\bibfnamefont {F.}~\bibnamefont {Haupt}},
  \bibinfo {author} {\bibfnamefont {L.~R.}\ \bibnamefont {Schreiber}}, \bibinfo
  {author} {\bibfnamefont {B.~E.}\ \bibnamefont {Kardyna{\l}}}, \ and\ \bibinfo
  {author} {\bibfnamefont {H.}~\bibnamefont {Bluhm}},\ }\href
  {https://link.aps.org/doi/10.1103/PhysRevB.99.205415} {\bibfield  {journal}
  {\bibinfo  {journal} {Physical Review B}\ }\textbf {\bibinfo {volume} {99}},\
  \bibinfo {pages} {205415} (\bibinfo {year} {2019})}\BibitemShut {NoStop}%
\bibitem [{\citenamefont {Havel}(2003)}]{Havel2003}%
  \BibitemOpen
  \bibfield  {author} {\bibinfo {author} {\bibfnamefont {T.~F.}\ \bibnamefont
  {Havel}},\ }\href {\doibase 10.1063/1.1518555} {\bibfield  {journal}
  {\bibinfo  {journal} {J. Math. Phys.}\ }\textbf {\bibinfo {volume} {44}},\
  \bibinfo {pages} {534} (\bibinfo {year} {2003})}\BibitemShut {NoStop}%
\end{thebibliography}
\end{document}